\documentclass[preprint, showpacs,preprintnumbers,amsmath,amssymb,nofootinbib]{revtex4}

 \usepackage{epsf}
 \usepackage{graphicx}
\usepackage{float}
\begin{document}
\newcommand{\be}[1]{\begin{equation}\label{#1}}
 \newcommand{\ee}{\end{equation}}
 \newcommand{\bea}{\begin{eqnarray}}
 \newcommand{\eea}{\end{eqnarray}}
 \def\disp{\displaystyle}

 \def\gsim{ \lower .75ex \hbox{$\sim$} \llap{\raise .27ex \hbox{$>$}} }
 \def\lsim{ \lower .75ex \hbox{$\sim$} \llap{\raise .27ex \hbox{$<$}} }

\title{\Large \bf Probing the course of cosmic expansion with a combination of observational data }
\author{Zhengxiang Li$^{1}$, Puxun Wu$^{2}$  and Hongwei Yu$^{1, 2} \footnote{Corresponding author: hwyu@hunnu.edu.cn}$ }

\address{$^1$Department of Physics and Key Laboratory of
Low Dimensional Quantum Structures and Quantum Control of Ministry
of Education, Hunan Normal University, Changsha, Hunan 410081, China\\
$^2$ Center for Nonlinear Science and Department of Physics, Ningbo
University,  Ningbo, Zhejiang 315211, China }

\begin{abstract}
We study the cosmic expansion history by reconstructing the
deceleration parameter $q(z)$ from the SDSS-II type Ia supernova
sample (SNIa) with two different light curve fits (MLCS2k2 and
SALT-II), the baryon acoustic oscillation (BAO) distance ratio, the
cosmic microwave background (CMB) shift parameter, and the lookback
time-redshift (LT) from the age of old passive galaxies. Three
parametrization forms for the equation of state of dark energy (CPL,
JBP, and UIS) are considered. Our results show that, for the CPL and
the UIS forms, MLCS2k2 SDSS-II SNIa+BAO+CMB and  MLCS2k2 SDSS-II
SNIa+BAO+CMB+LT favor a currently slowing-down  cosmic acceleration,
but this does not occur for all other cases, where an increasing
cosmic acceleration is still favored. Thus, the reconstructed
evolutionary behaviors of dark energy and the course of the cosmic
acceleration are highly dependent both on the light curve fitting
method for the SNIa and the parametrization form for the equation of
state of dark energy.
\end{abstract}

\pacs{95.36.+x, 98.80.Es}

 \maketitle
 \renewcommand{\baselinestretch}{1.5}

\section{INTRODUCTION}\label{sec1}
One of the most important and mysterious issues of modern cosmology
is the fact that our universe is undergoing an accelerating
expansion~\cite{expansion1, expansion2}.  This observed phenomenon
 is usually attributed to the existence of  an exotic energy
component called dark energy (DE), which generates a repulsive force
due to the negative pressure associated with it. The simplest
candidate of DE is the cosmological constant with the equation of
state (EOS) $w=-1$.  If one generalizes the EOS  $w$ of DE from
$w=-1$ to be an arbitrary constant $w$, the current astronomical
observations~\cite{observation1, observation2, observation3} show
that $w$ is confined to be $|1+w|<0.06$ at the $1\sigma$ confidence
level. However,  the EOS might also be a function
 of cosmic time. In order to unveil the evolutionary properties of dark
energy using observations, one usually adopts a parametrization form
with some free parameters for $w(z)$, which may not be motivated by
any particular foundamental theory. Examples of such kind  are the
Chevallier-Polarski-Linder (CPL) parametrization~\cite{CPL}, the
Jassal-Bagla-Padmanabhan (JBP) parametrization~\cite{JBP} as well as
the Upadhye-Ishak-Steinhardt (UIS) parametrization~\cite{UIS}, and
so on. Constraining the free parameters of a given parametrization
with observational data, one can obtain the evolutionary curve of
$w(z)$, which embodies the property of dark energy. For instance,
some current observations give an indication that the EOS has
crossed the phantom divider ($w=-1$) at least once~\cite{cross1,
cross2}.

Recently, Sahni et al. proposed a new diagnostic of DE, named
$Om(z)$ diagnostic. The advantage of this diagnostic, as opposed to
the EOS, is that it depends on the first derivative of the
luminosity distance $d_L(z)$~\cite{diagnostic}, and thus is less
sensitive to the observational errors and the present matter energy
density $\Omega_{0m}$.  In addition, one can discriminate DE models
with the EOS $w<-1, w=-1$ and $w>-1$ effectively by examining the
slope of $Om(z)$  even if the value of $\Omega_{0m}$ is not exactly
known, with positive, null, or negative slopes corresponding to
$w<-1, w=-1$ or $w>-1$, respectively.

Performing the $Om(z)$ diagnostic~\cite{diagnostic} and analyzing
the deceleration parameter $q(z)$ with the Constitution type Ia
supernova data (SNIa)~\cite{constitution} and data from the baryon
acoustic oscillation (BAO) distance ratio of the distance
measurements obtained at $z = 0.2$ and $z = 0.35$ in the galaxy
power spectrum~\cite{bao1, bao2} by using the popular CPL
parametrization, Shafieloo et al.  found that $Om(z)$  and $q(z)$
increases markedly at the low redshifts $z< 0.3$~\cite{slowing}.
This result suggests that the expansion acceleration of our universe
is probably slowing down. However, the result obtained from a
combination of the cosmic microwave background (CMB) shift
parameter,  SNIa and BAO is very well consistent with the
$\Lambda$CDM model. So, there appears some tension between low
redshift data (Constitution SNIa+BAO) and high redshift (CMB) one.
Surprisingly, further analysis using a subsample (SNLS+ESSENCE+CfA)
of the Constitution SNIa reveals that the outcome does not rely on
whether the CMB data is added and the cosmic acceleration has been
over the peak. It was therefore argued the tension  could be either
due to the systematics in some data or that the CPL parametrization
is not versatile-enough to accommodate the evolution of DE implied
by the data. This situation was also examined  by Gong {\it et
al.}~\cite{gong} recently through the reconstruction of $Om(z)$.
They found that both the systematics in data sets and the
parametrization of DE influence the evolutional behavior of DE. It
is worth noting that the results in Ref.~\cite{slowing} are obtained
only at the $1\sigma$ confidence level and whether they are reliable
at the $2\sigma$ confidence level still needs to be checked.

In this paper, we will re-examine this issue with three different
parametrization forms for the EOS of DE (CPL, JBP and UIS~\cite{CPL,
JBP, UIS}), and  288 SNIa data points released by the Sloan Digital
Sky Survey-II (SDSS-II) Supernova Survey with two different light
curve fits (MLCS2k2 fit and SALT-II fit).  We hope this will help us
further understand the influence of different light curve fitting
methods~\footnote{The influence of different light curve fits has
been considered in \cite{gong}. However, there, different data
points are obtained with different light curve fits for Constitution
SNIa. In the present paper, data points are the same for different
light curve fits.}. As in~\cite{slowing}, the BAO distance ratio and
the CMB shift parameter are also considered in our analysis. In
addition, we use the lookback time-redshift (LT) from the age of old
passive galaxies, since it has the advantage that the ages of
distant objects are independent of each other, and thus  may avoid
biases that are present in techniques using distances of primary or
secondary indicators in the cosmic distance ladder method.

\section{PARAMETRIZATION}\label{sec2}
By assuming that the energy components of our universe are
nonrelativistic matter and dark energy, the Friedmann equation can
be expressed as
\begin{equation}\label{E2}
E^2(z)=\frac{H^2(z)}{H_0^2}=\Omega_{0m}(1+z)^3+\Omega_{DE},
\end{equation}
where $\Omega_{0m}$ is the current value of the dimensionless matter energy density,
and $\Omega_{DE}$ is the dimensionless energy density parameter of dark energy,
which can be expressed as
\begin{equation}\label{om}
\Omega_{DE}=(1-\Omega_{0m})\exp\bigg[3\int_0^z\frac{1+w(z')}{1+z'}dz'\bigg]\;,
\end{equation}
where $w(z)$ is the equation of state of dark energy.

Now, we consider three parametrization forms for the EOS. i.e.,  the
CPL, JBP and UIS parametrization. The EOS $w(z)$ for the CPL
is~\cite{CPL}
\begin{equation}
w(z)=w_0+\frac{w_1z}{1+z}.
\end{equation}
Substituting it into Eqs.~(\ref{E2}, \ref{om}), one has
\begin{equation}
E^2(z)=
\Omega_{0m}(1+z)^3+(1-\Omega_{0m})(1+z)^{3(1+w_0+w_1)}\exp\bigg(-\frac{3w_1z}{1+z}\bigg)\;.
\end{equation}
For the JBP parametrization,  $w(z)$ is~\cite{JBP}
\begin{equation}
w(z)=w_0+\frac{w_1z}{(1+z)^2}\;.
\end{equation}
So we have
\begin{equation}
E^2(z)= \Omega_{0m}(1+z)^3+(1-\Omega_{0m})(1+z)^{3(1+w_0)}\exp\bigg(\frac{3w_1z^2}{2(1+z)^2}\bigg)\;.
\end{equation}
And for the UIS parametrization~\cite{UIS}, the $w(z)$ and $E^2(z)$
can be expressed  respectively as,
\begin{equation}
 w(z)=\begin{cases} w_0+w_1z ~~&\mbox{$z<1$},\\w_0+w_1 ~~&\mbox{$z\geq1$,}
\end{cases}
\end{equation}
and
\begin{equation}
 E^2(z)=\begin{cases} \Omega_{0m}(1+z)^3+(1-\Omega_{0m})(1+z)^{3(1+w_0-w_1)}\exp(3w_1z)
 ~~&\mbox{$z<1$},\\ \Omega_{0m}(1+z)^3+(1-\Omega_{0m})(1+z)^{3(1+w_0+w_1)} ~~&\mbox{$z\geq1$.}
\end{cases}
\end{equation}

\section{OBSERVATIONAL DATA}\label{sec3}
The SNIa dataset used in our analysis is the 288 data points
released by the   Sloan Digital Sky Survey-II (SDSS-II) Supernova
Survey~\cite{Kessler}, which  consist of  $103$ new SNIa from the
first-year SDSS-II supernova survey~\cite{Frieman,Sako,Kessler}, 56
points from ESSENCE~\cite{Wood-Vasey}, 62 from SNLS~\cite{Astier},
34 from HST~\cite{Riess}, and 33 nearby SNIa~\cite{Jha}. For 288
SDSS-II SNIa data,  two kinds of light curve fitting methods, i.e.,
the MLCS2k2 fit and the SALT-II fit, have been
employed~\cite{Kessler}. Here, for the sake of systematics influence
check, we do our analysis of the SDSS-II data with both  kinds of
fits.

The constraints from the SNIa data can be obtained by minimizing the
following $\chi^2$ statistics:
\begin{equation}\label{Chi2}
\chi_\mu^2(\mu_0, \textbf{p})=\sum_{i=1}^{288}\frac{[\mu_{obs,
 i}-\mu_{th}(z_i;\mu_0,\textbf{p})]^2}{\sigma_\mu^2},
\end{equation}
where $\mu_{obs,i}$ is the distance modulus estimated from the
MLCS2K2 fit or SALT-II fit for the i'th supernova, $z_i$ is its
spectroscopically determined redshift, $\textbf{p}$ stands for the
complete set of model parameters, and $\mu_{th}(z_i;\textbf{p})$ is
the theoretical distance modulus for a concrete cosmological model,
which is calculated from
\begin{equation}
\mu_{th}(z_i;\mu_0,
\textbf{p})=5\log_{10}(d_L(z_i;\textbf{p}))+\mu_0,
\end{equation}
where $d_L$ represents the luminosity distance. For a flat universe,
$d_L$ is given by
\begin{equation}
d_L(z;\textbf{p})=(1+z)\int_0^zdz'\frac{1}{E(z', \textbf{p})}.
\end{equation}
Let us note that here  $\mu_0$ is a nuisance parameter.
In order to marginalize over it, we expand
$\chi_\mu^2$ (Eq. (\ref{Chi2})) with respect to $\mu_0$ as
\begin{equation}\label{chi22}
\chi_\mu^2(\mu_0, \textbf{p})=A-2\mu_0B+\mu_0^2C\;,
\end{equation}
where
\begin{equation}
A(\textbf{p})=\sum_{i=1}^{288}\frac{[\mu_{obs,i}-\mu_{th}(z_i;\mu_0=0,\textbf{p})]^2}{\sigma_\mu^2}\;,
\end{equation}
\begin{eqnarray}
B(\textbf{p})&=&\sum_{i=1}^{288}\frac{[\mu_{obs,i}-\mu_{th}(z_i;\mu_0=0,\textbf{p})]}{\sigma_\mu^2}\;,\\
C&=&\sum_{i=1}^{288}\frac{1}{\sigma_\mu^2}\;.
\end{eqnarray}
Eq.~(\ref{chi22}) has a minimum at $\mu_0=B/C$, and it is
\begin{equation}
\tilde{\chi}_\mu^2(\textbf{p})=A(\textbf{p})-\frac{B(\textbf{p})^2}{C}\;.
\end{equation}
Thus, instead of minimizing $\chi_\mu^2(\mu_0,\textbf{p})$, we can
minimize $\tilde{\chi}_\mu^2(\textbf{p})$, which is independent of
$\mu_0$, to obtain  constraints on the model parameters.

In the $\chi_\mu^2$-expression, the distance-modulus uncertainty is
given by
\begin{equation}
\sigma_\mu^2=(\sigma_\mu^{fit})^2+(\sigma_\mu^{int})^2+(\sigma_\mu^z)^2,
\end{equation}
where $\sigma_\mu^{fit}$ and $\sigma_\mu^{int}$(=0.16 for MLCS2k2
and 0.14 for SALT-II) are  respectively the statistical uncertainty
and the additional (intrinsic) error derived from light-curve
fitting method.   $\sigma_\mu^{int}$ is introduced to unitize
$\chi^2$ per degree of freedom for the Hubble diagram constructed
from the nearby SNIa sample and it is detailed in
Ref.~\cite{Kessler}. It has  little effect on the results of model
parameters, but improves the value of $\chi^2_\mu$ remarkably. For
example, for SALT-II SNIa with and without $\sigma_\mu^{int}$, the
best fit values of the CPL parametrization  are $w_0=-1.022$ and
$w_1=0.090$ with $\chi^2_\mu=268.66$, and $w_0=-1.015$ and
$w_1=0.200$ with $\chi^2_\mu=589.31$, respectively, where
$\Omega_{0m}$ is given prior to be $0.262$ obtained from
WMAP7~\cite{WMAP7}. $\sigma_\mu^z$ is the error which relates with
the redshift uncertainty and can be computed from
\begin{equation}
\sigma_\mu^z=\sigma_z\bigg(\frac{5}{\ln10}\bigg)\frac{1+z}{z(1+z/2)},
\end{equation}
where $\sigma_z^2$ is the redshift uncertainty and is defined to be
\begin{equation}
\sigma_z^2=\sigma_{z,spec}^2+\sigma_{z,pec}^2.
\end{equation}
Here, $\sigma_{z,spec}$ and $\sigma_{z,pec}$, as discussed in detail
in Ref.~\cite{Kessler}, are the uncertainties from spectroscopic
measurements and peculiar motion of the host galaxy, respectively.

The BAO distance measurements used in our analysis are obtained at
$z=0.2$ and $z=0.35$ from the joint analysis of the 2dF Galaxy
Redshift Survey and SDSS data~\cite{bao2}. The BAO distance ratio
\begin{equation}
\frac{D_V(z=0.35)}{D_V(z=0.20)}=1.736\pm 0.065
\end{equation}
is a relatively model-independent quantity. Here $D_V(z)$ is defined
as
\begin{equation}
D_V(z_{BAO})=\bigg[\frac{z_{BAO}}{H(z_{BAO})}\bigg(\int_0^{z_{BAO}}\frac{dz}{H(z)}\bigg)^2\bigg]^{1/3}.
\end{equation}
For the BAO dataset, we can fit the model parameter $\mathbf{p}$ by
performing $\chi^2$ statistics as follows
\begin{equation}
\chi_{BAO}^2(\mathbf{p})=\frac{[D_V(z=0.35)/D_V(z=0.20)-1.736]^2}{0.065^2}\;.
\end{equation}

In addition  to the low redshift SNIa and BAO, we add the high
redshift CMB parameter which is the reduce distance at
$z_{ls}=1090$. The shift parameter
\begin{equation}
R(\textbf{p})=\sqrt{\Omega_{0m}}\int_0^{z_{ls}}\frac{dz}{E(z,\textbf{p})}=1.71\pm 0.019,
\end{equation}
 is used. We also apply  $\chi^2$
\begin{equation}
\chi^2_{CMB}(\mathbf{p})=\frac{[R(\mathbf{p})-1.71]^2}{0.019^2}
\end{equation}
to the model parameter $\mathbf{p}$ for the CMB data.

Besides the most common observational data sets above, we also
perform an  analysis combined with the lookback time-redshift data
(LT), which is established by estimating the age of 32 old passive
galaxies distributed over the redshift interval $0.11\leq z\leq1.84$
and the total age of the universe $t_0^{obs}$~\cite{Simon}. The
galaxy samples of passively evolving galaxies are selected with
high-quality spectroscopy and the method used to determine ages of
galaxy samples indicates that systematics are not a serious source
of error for these high-redshift galaxies. In addition, this data
set has the advantage that the ages of distant objects are
independent of each other, and thus it may avoid biases that are
present in techniques that use distances of primary or secondary
indicators in the cosmic distance ladder method. As a result, these
age data are different from the widely used distance one, such as
SNIa, and it may help us gain more insight into the nature of dark
energy. To estimate the best fit of model parameters, we minimize
$\chi_{age}^2$
 \begin{equation}\label{Chi2age}
\chi_{age}^2(\mathbf{p})=\sum\limits_i
\frac{\big[t_L(z_i;\textbf{p})-t_L^{obs}(z_i;\tau)\big]^2}{\sigma_{T}^2}+\frac{\big[t_0(\textbf{p})-t_0^{obs}\big]}{\sigma_{t_o^{obs}}^2},
 \end{equation}
where, $\sigma_T^2\equiv\sigma_i^2+\sigma_{t_0^{obs}}^2$, $\sigma_i$
is the uncertainty in the individual lookback time to the $i^{th}$
galaxy of the sample,  $\sigma_{t_0^{obs}}=0.7~Gyr$ stands for the
uncertainty in the total expansion age of the universe
($t_0^{obs}$), and $\tau$ means the time from Big Bang to the
formation of the object, which is the so-called delay factor or
incubation time. Note that while the observed lookback time
($t_L^{obs}(z_i;\tau)$) is directly dependent on $\tau$, its
theoretical value ($t_L(z_i; \textbf{p})$) is not. Furthermore,  in
principle, it must be different for each object in the sample. Thus
the delay factor becomes  a ``nuisance" parameter,  we use the
following method to marginalize over it~\cite{obl, modlike}
 \begin{eqnarray}
\tilde{\chi}^2 (\mathbf{p})&=&-2\ln\int_0^\infty
d\tau\exp\bigg(-\frac{1}{2}\chi_{age}^2\bigg)\nonumber\\
&=&A-\frac{B^2}{C}+D-2\ln\bigg[\sqrt{\frac{\pi}{2C}}\textrm{erfc}\bigg(\frac{B}{\sqrt{2C}}\bigg)\bigg],
 \end{eqnarray}
where
 \be{eq8}
A=\sum_{i=1}^n
\frac{\triangle^2}{\sigma_{T}^2}.~~~~~~~~~~~~~~B=\sum_{i=1}^n\frac{\triangle}{\sigma_{T}^2}.~~~~~~~~~~~~~~C=\sum_{i=1}^n\frac{1}{\sigma_{T}^2},
 \ee
 D is the second term of the rhs of Eq.~(\ref{Chi2age}),
 \begin{equation}
\triangle=t_L(z_i;\textbf{p})-[t_0^{obs}-t(z_i)],
 \end{equation}
and $\textrm{erfc}(x)$ is the complementary error function of the variable $x$.

\section{RESULTS}\label{sec4}
\begin{figure}[htbp]
 \centering
\includegraphics[width=0.45\textwidth, height=0.38\textwidth]{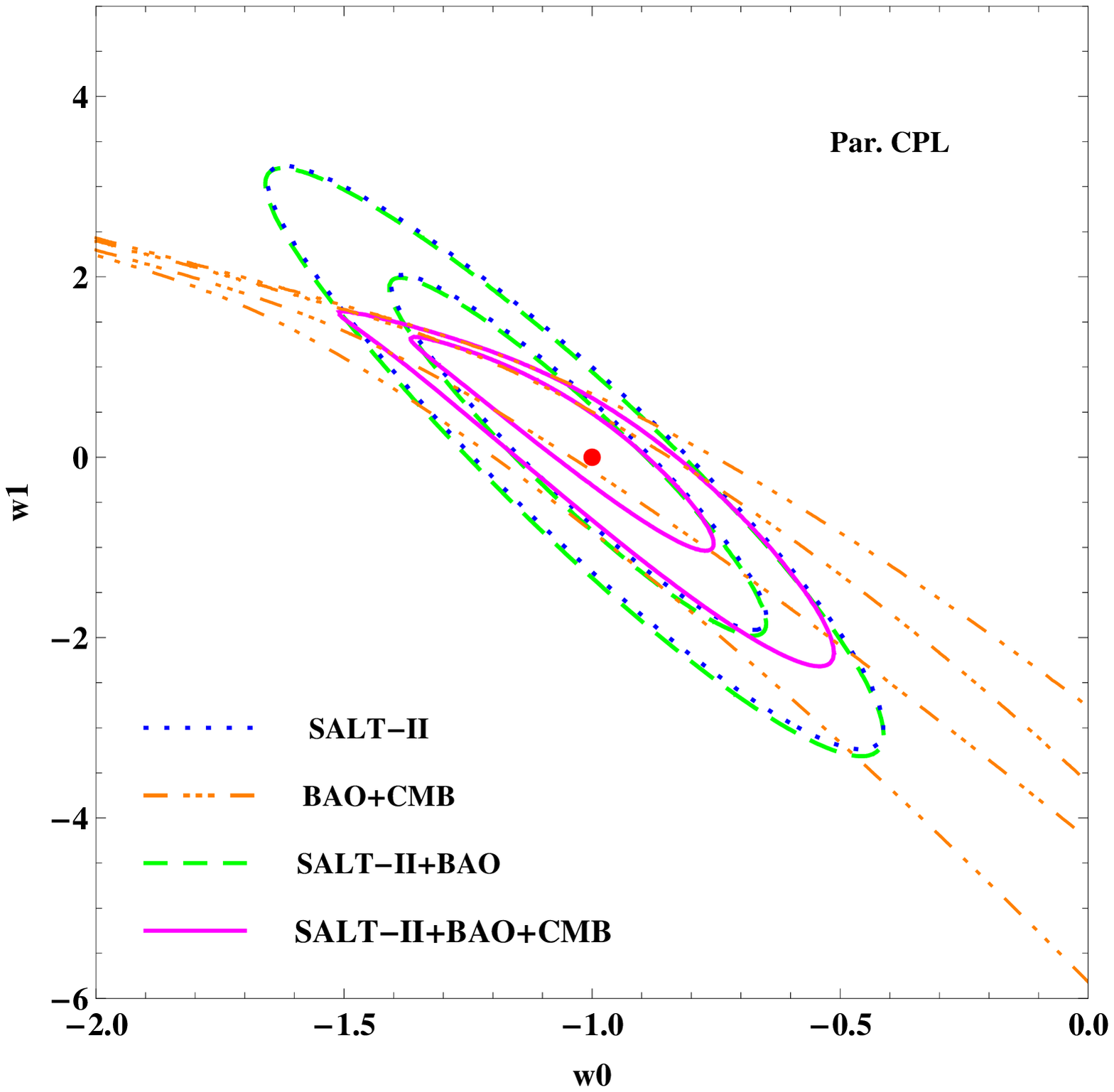}\includegraphics[width=0.45\textwidth, height=0.38\textwidth]{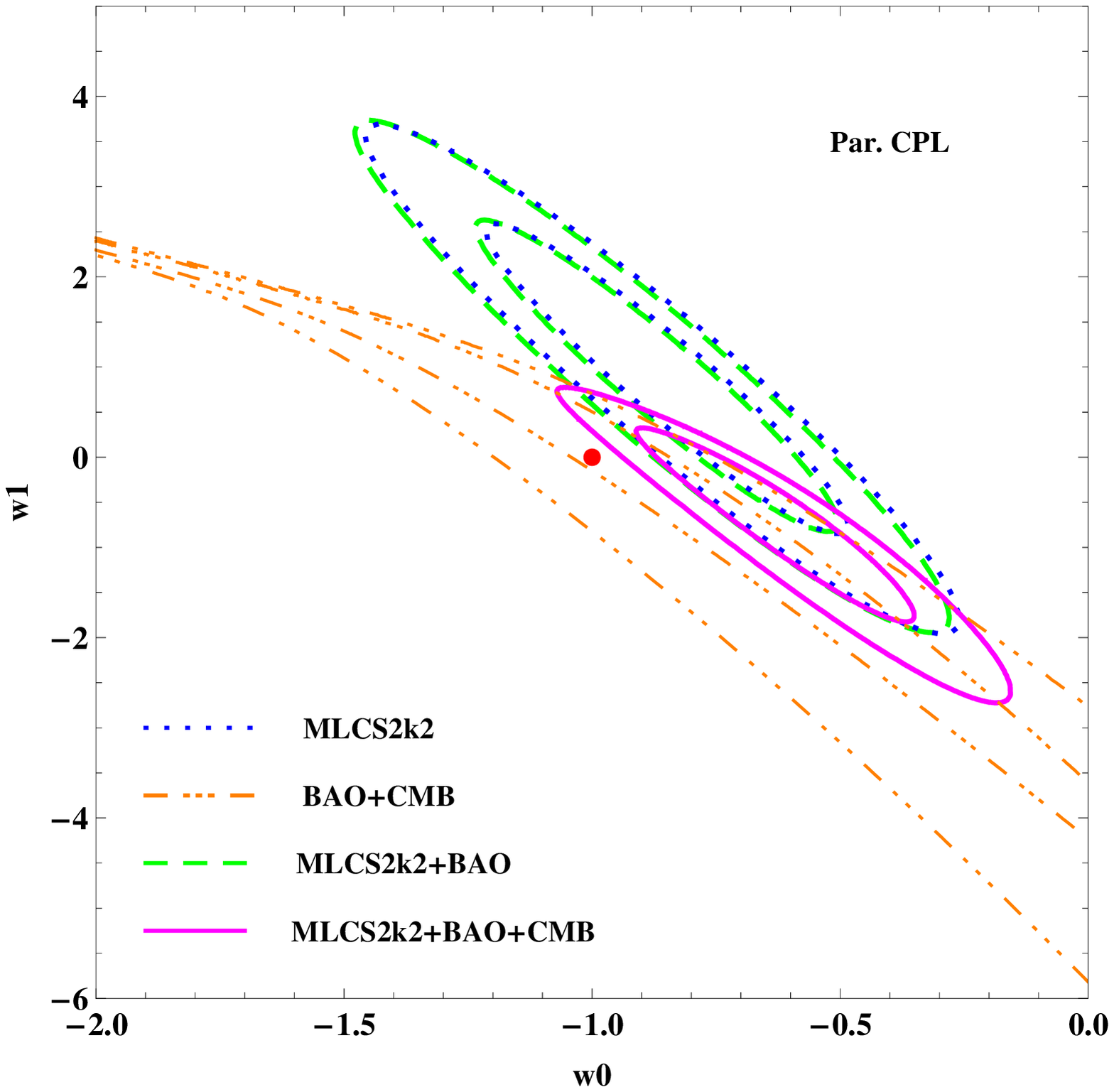}
\includegraphics[width=0.45\textwidth, height=0.38\textwidth]{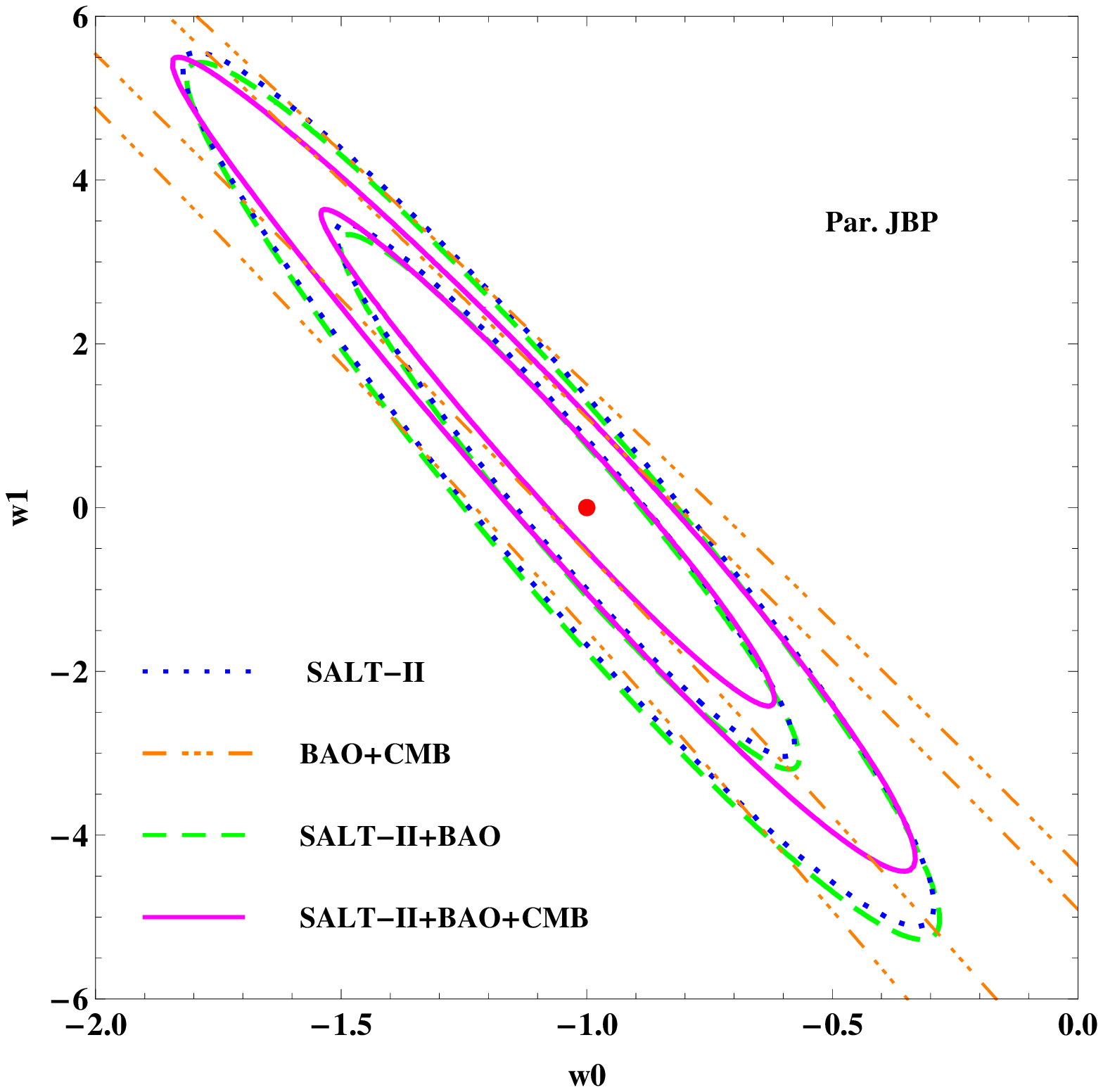}\includegraphics[width=0.45\textwidth, height=0.38\textwidth]{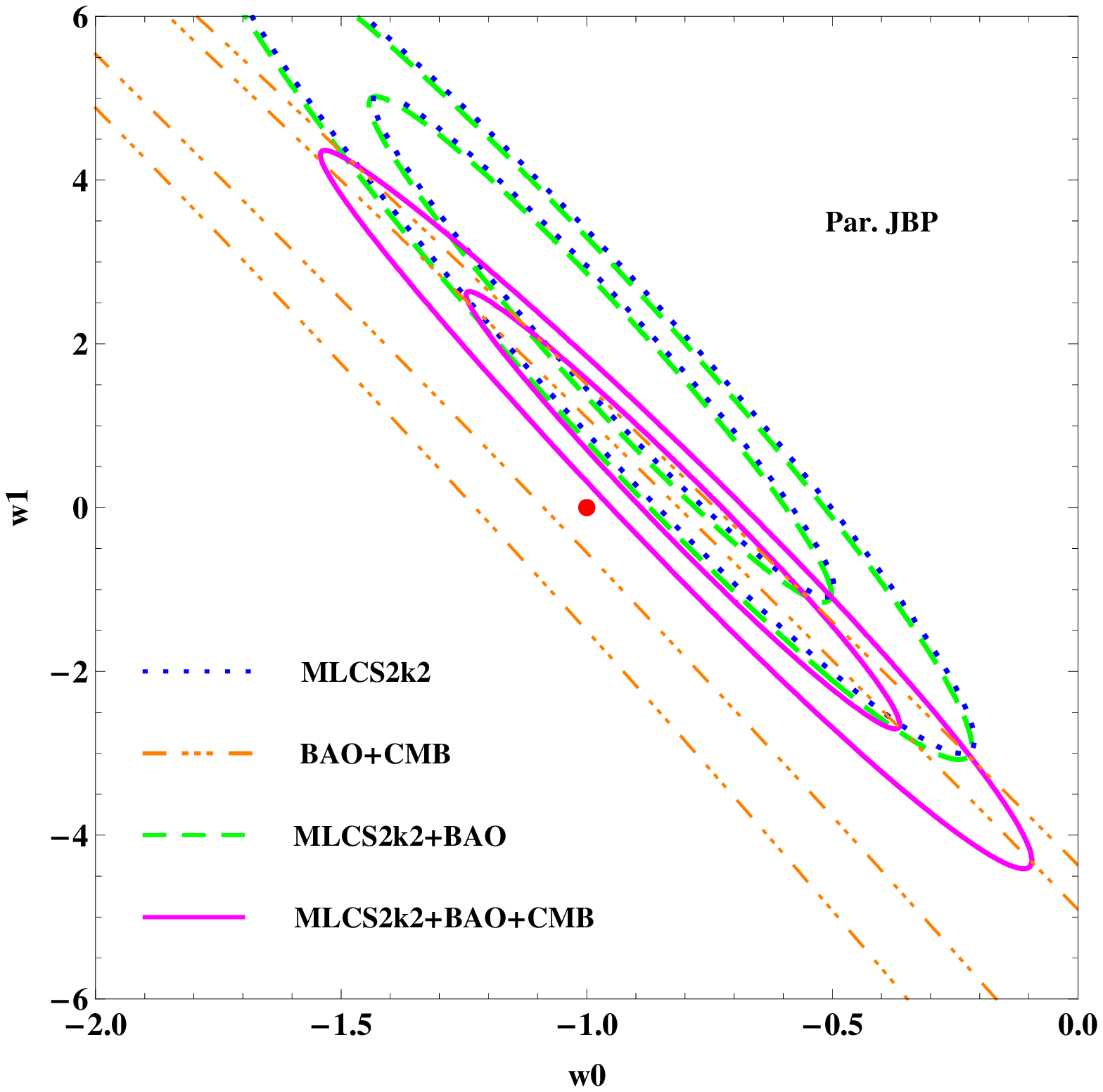}
\includegraphics[width=0.45\textwidth, height=0.38\textwidth]{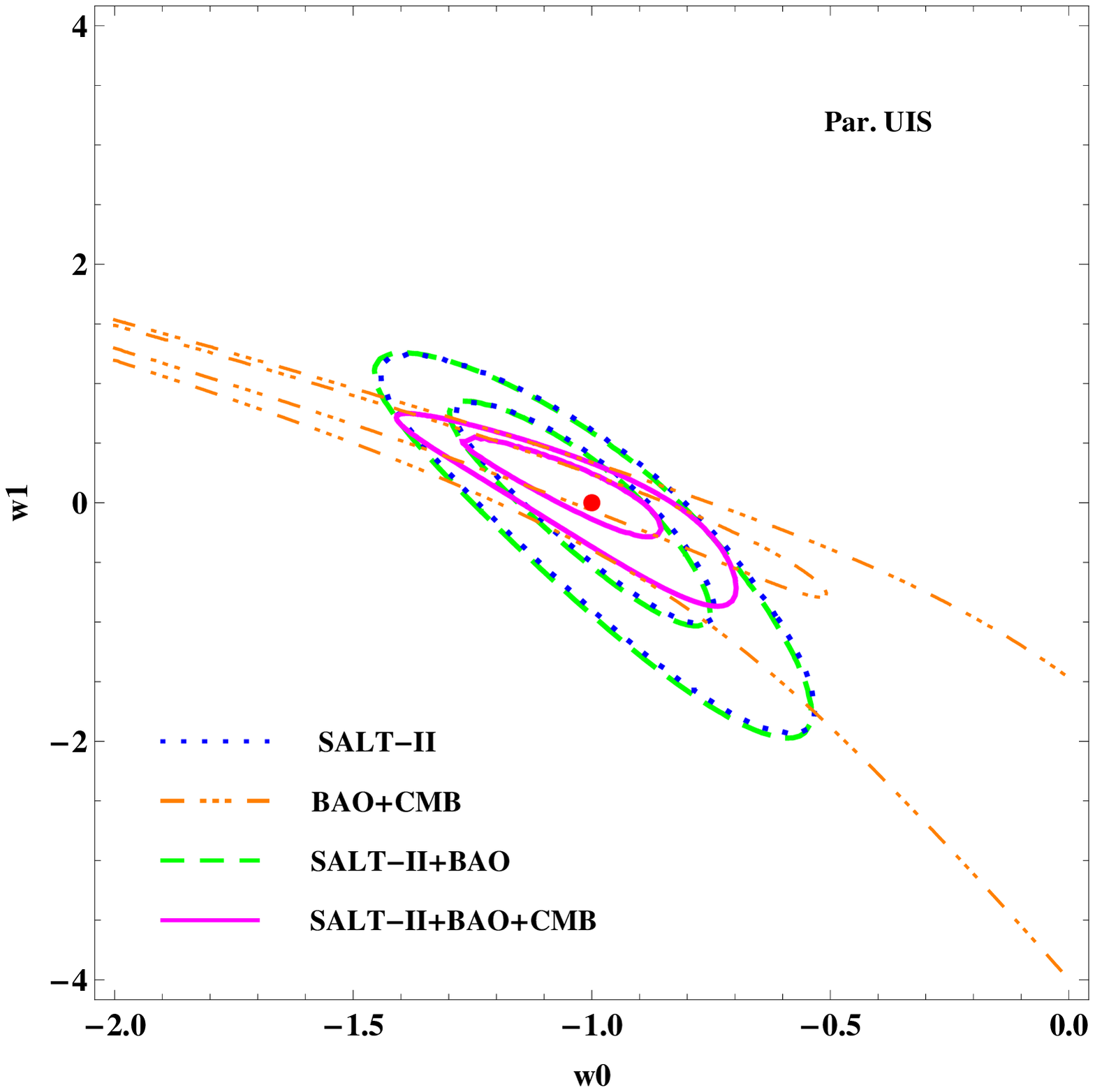}\includegraphics[width=0.45\textwidth, height=0.38\textwidth]{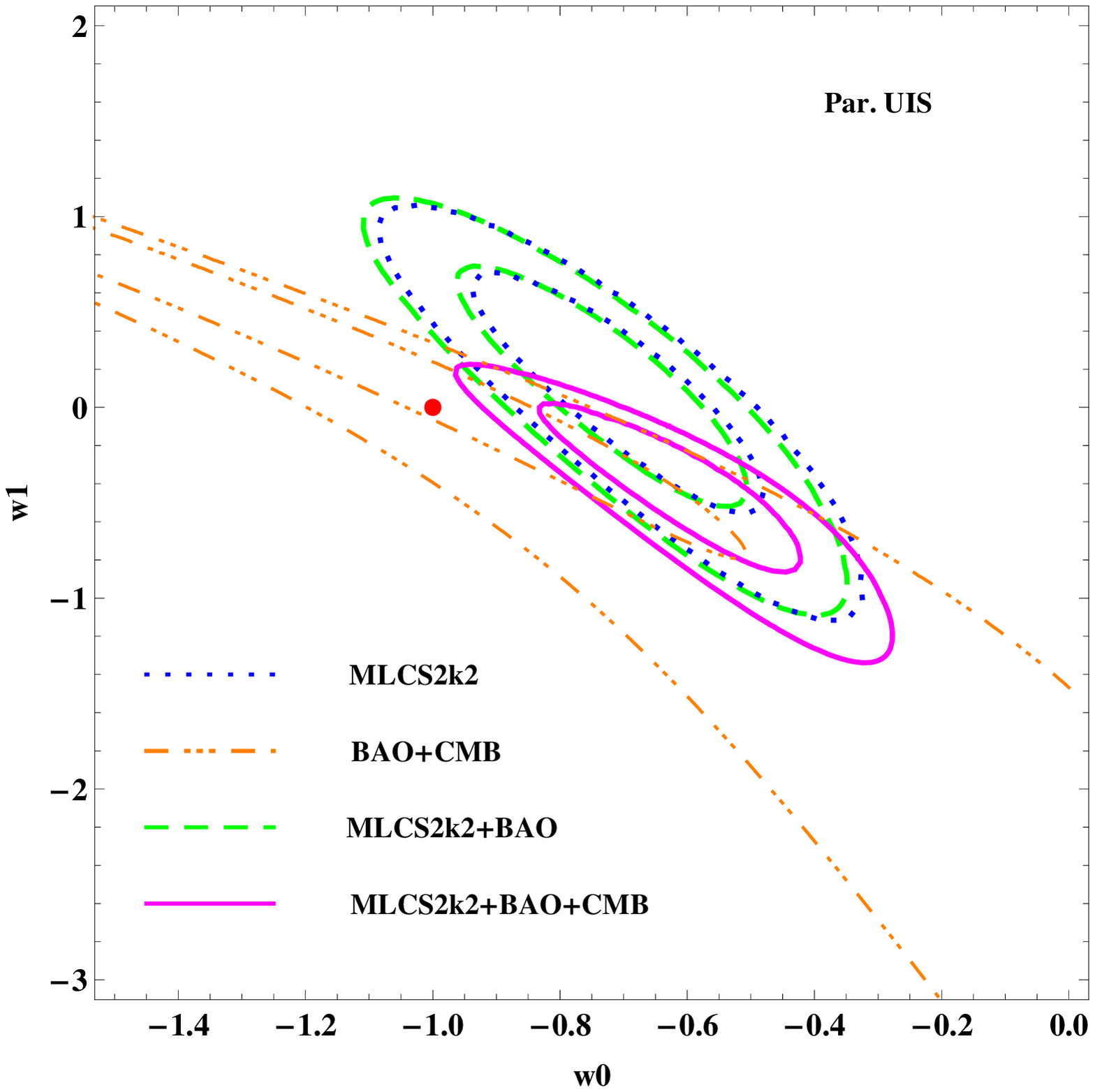}
 \caption{\label{Fig2}
The $68.3\%$ and $95\%$ confidence level regions
   for $w_0$ versus $w_1$. The blue dotted, yellow dot-dashed, green dashed and
   pink solid lines represent the results obtained from
SNIa, BAO+CMB, SNIa+BAO and SNIa+BAO+CMB, respectively.  The left
panels show the results obtained from SNIa with SALT fit, while the
right panels are the results from SNIa with MLCS2k2 fit.
   The red point at $w_0=-1$, $w_1=0$
   represents the spatially flat $\Lambda$CDM model.}
   \end{figure}

\begin{figure}[htbp]
 \centering
\includegraphics[width=0.45\textwidth, height=0.38\textwidth]{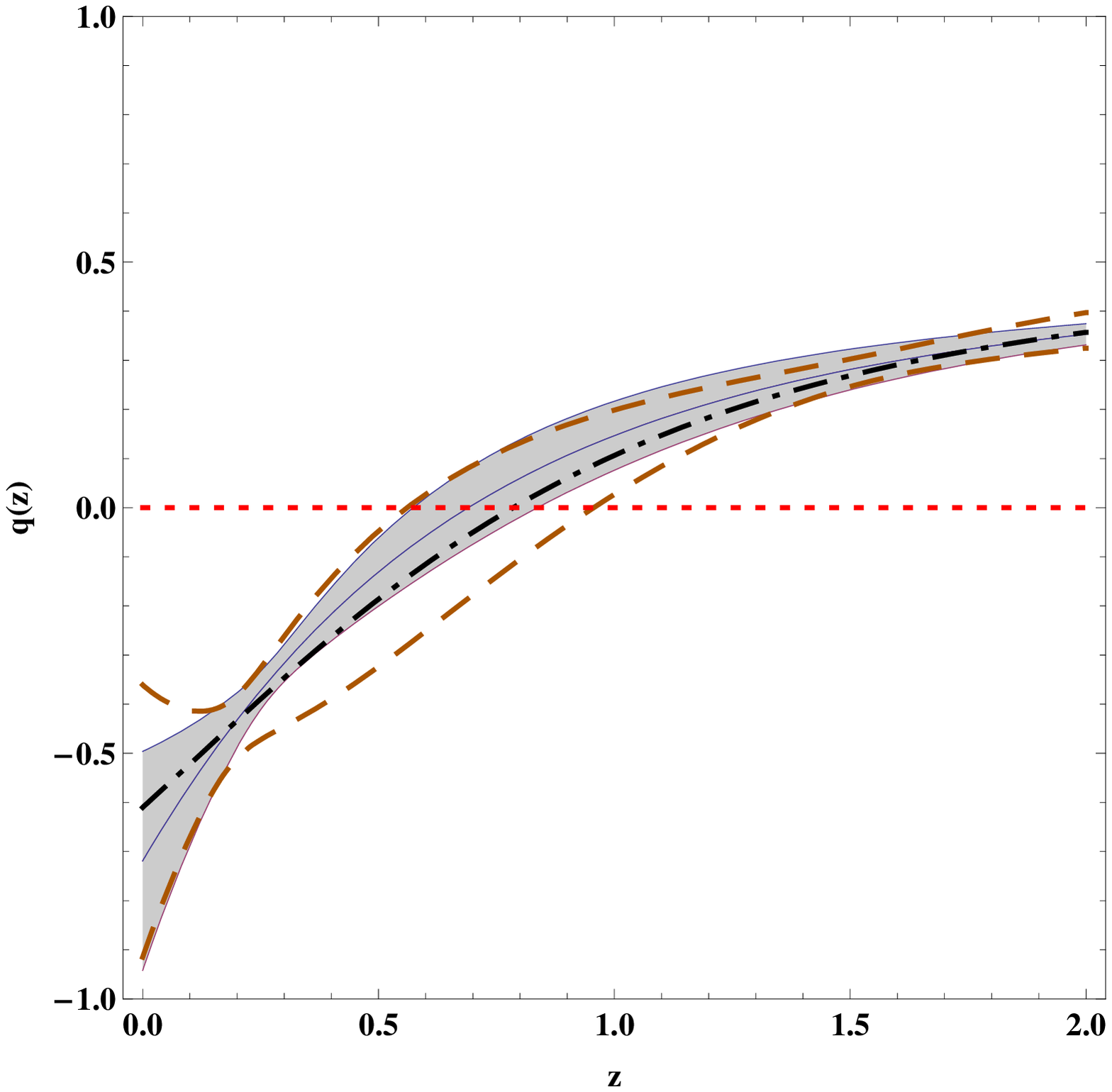}\includegraphics[width=0.45\textwidth, height=0.38\textwidth]{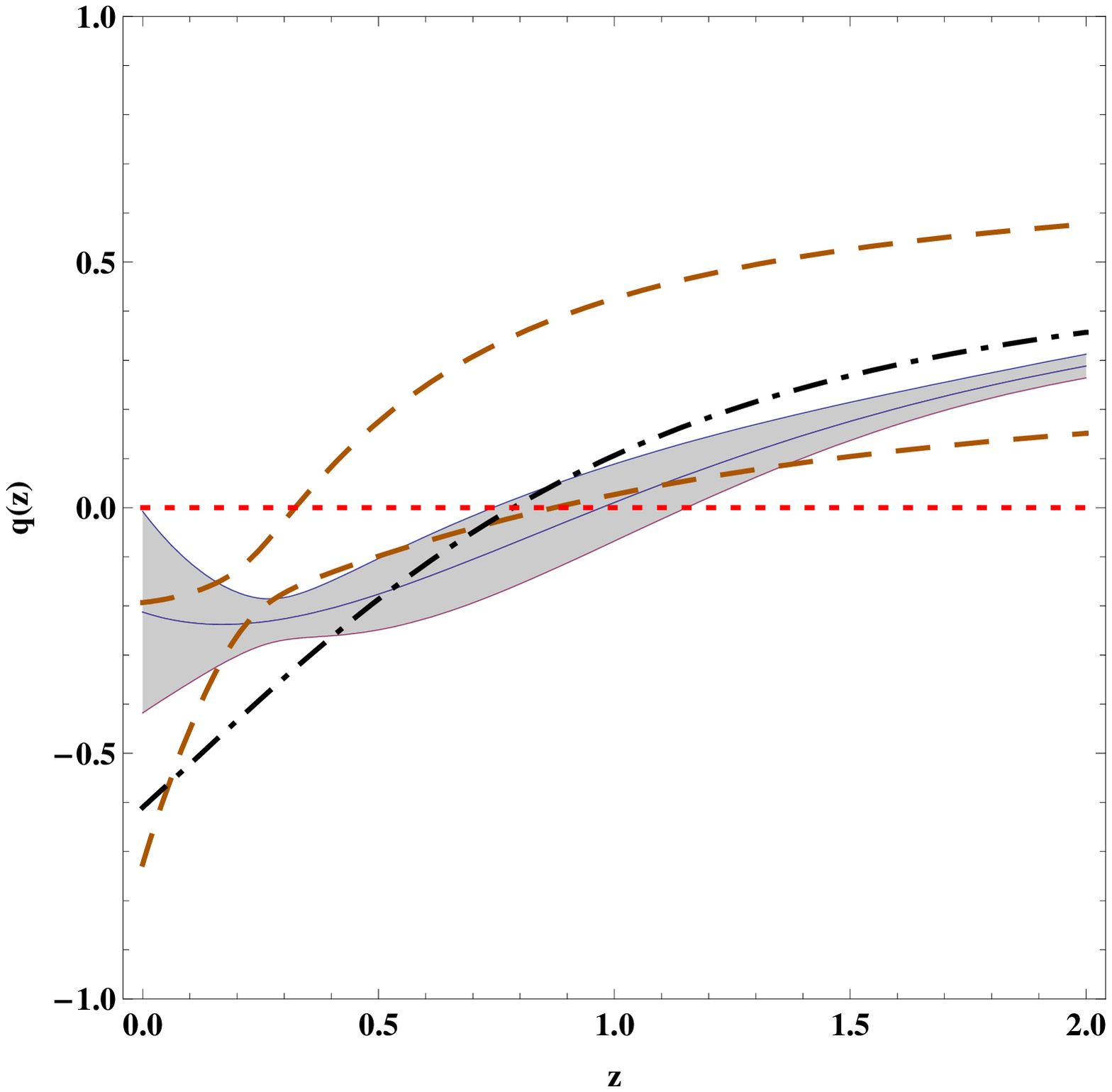}
\includegraphics[width=0.45\textwidth, height=0.38\textwidth]{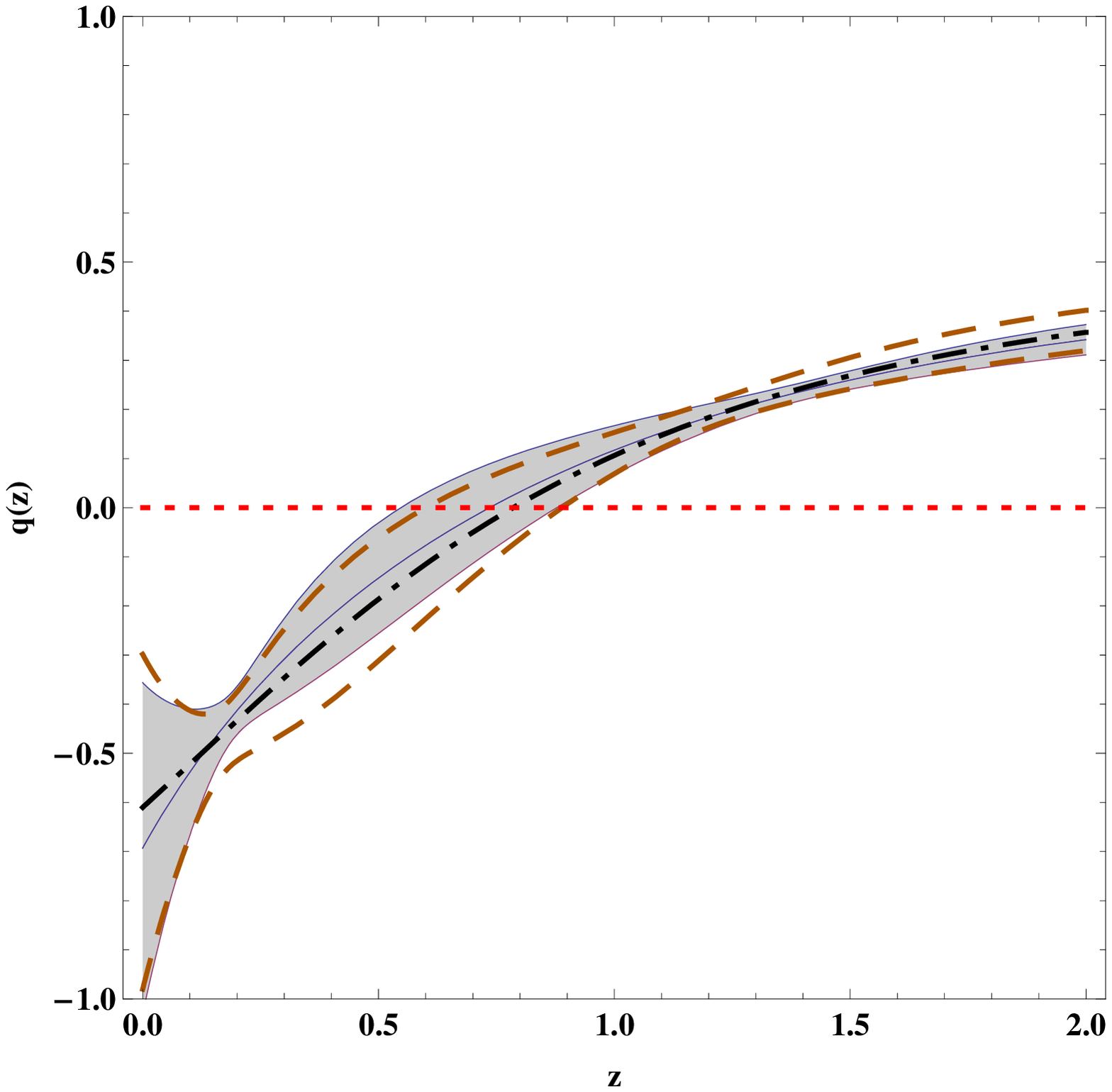}\includegraphics[width=0.45\textwidth, height=0.38\textwidth]{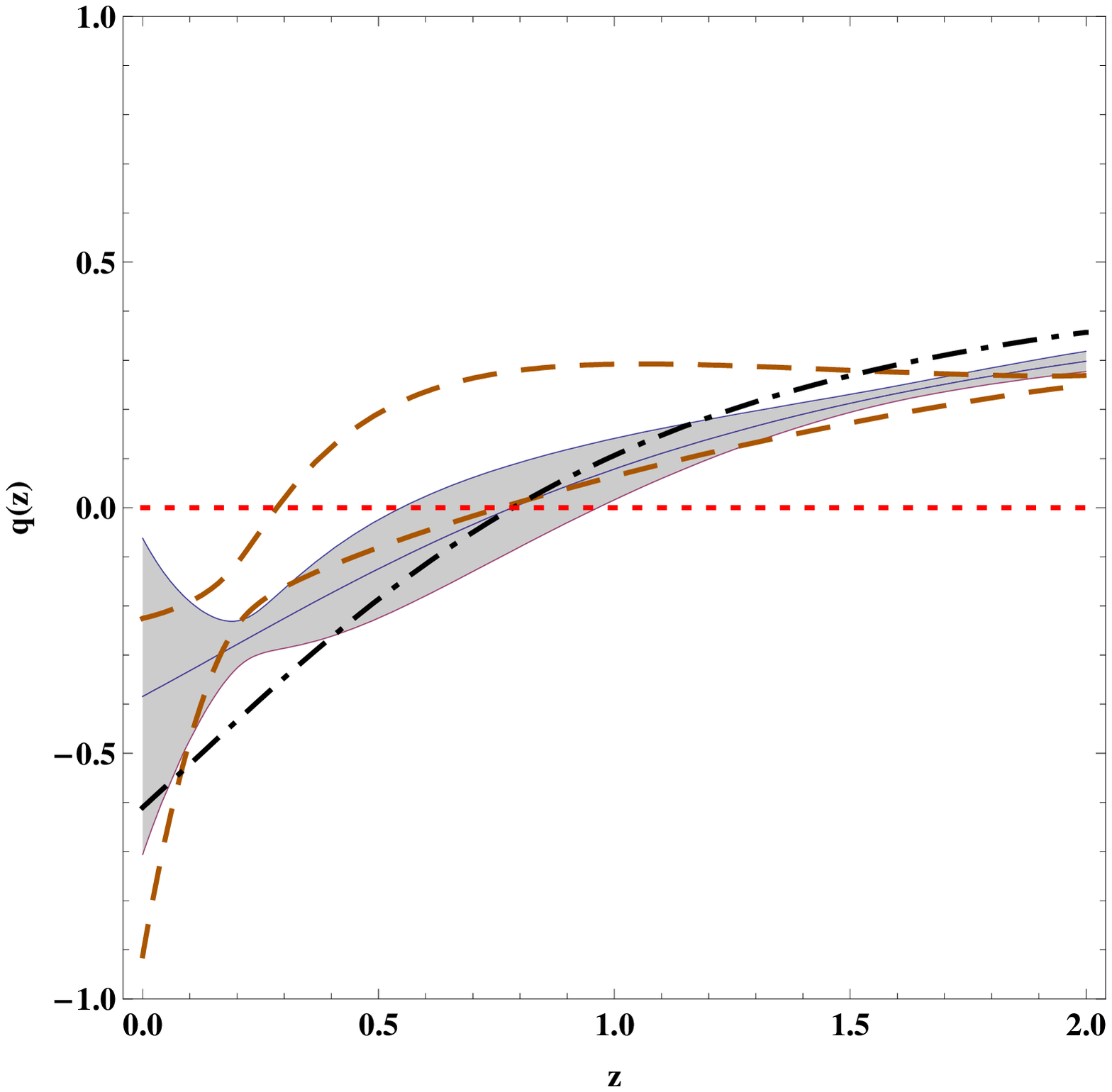}
\includegraphics[width=0.45\textwidth, height=0.38\textwidth]{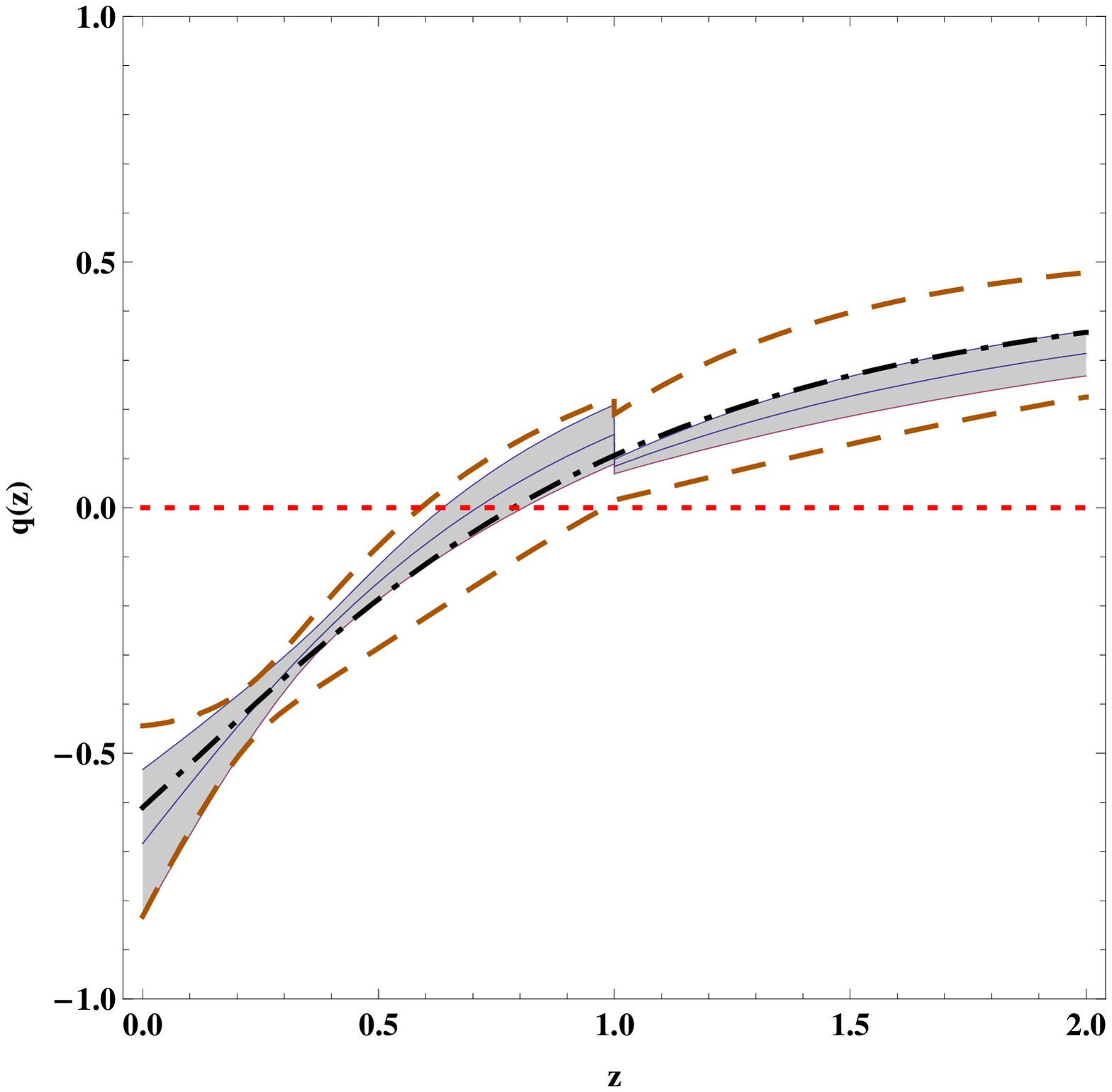}\includegraphics[width=0.45\textwidth, height=0.38\textwidth]{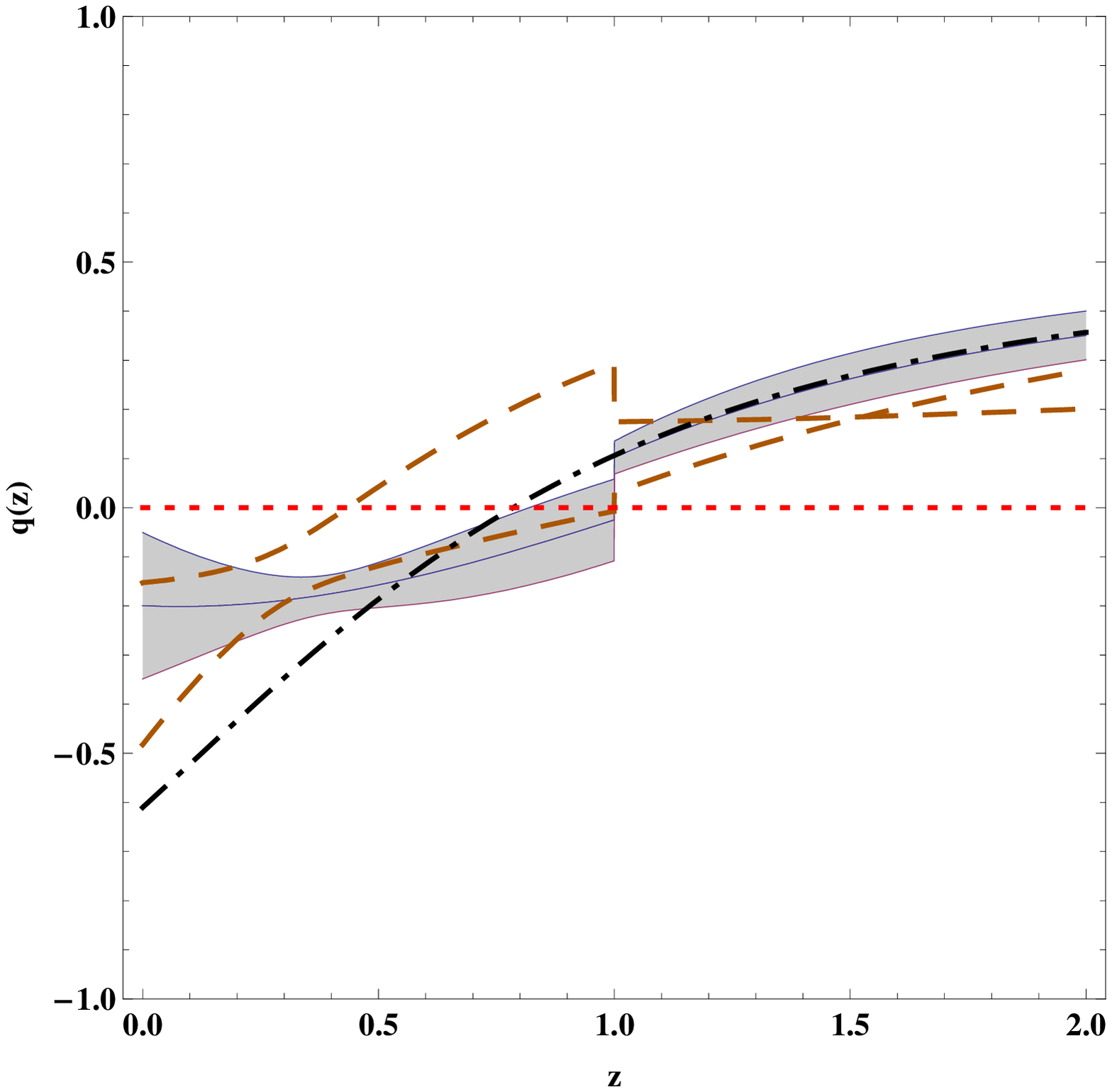}
\caption{\label{Fig3} The gray regions and the regions between two
dashed lines show the evolutionary behaviors of $q(z)$ at the
$68.3\%$ confidence level obtained from SNIa+BAO+CMB and SNIa+BAO,
respectively.   The left panels show the results obtained from SNIa
with SALT fit, while the right panels are the results from SNIa with
MLCS2k2 fit. The dot-dashed lines represent the best-fit spatially
flat $\Lambda$CDM model.  }
 \end{figure}

We first explore three popular parametrization forms by using
SDSS-II SNIa, BAO and CMB. A comparison of two light curve fits of
SDSS-II SNIa is given. The results are shown in Figs.~(\ref{Fig2},
\ref{Fig3}).

Fig.~(\ref{Fig2}) gives the contour diagrams on $w_0-w_1$ panel at
the $68.3\%$ and $95.4\%$ confidence levels for three
parametrization forms. Since different data sets give different best
fit values of $\Omega_{0m}$, we set prior
 $\Omega_{0m}=0.262$ in this paper, which is
the best fit value from the  WMAP7~\cite{WMAP7}. The top, middle,
and bottom panels are the results of the CPL, JBP  and UIS,
respectively. The big red points denote the spatially flat
$\Lambda$CDM model ($w_0=-1$ and $w_1=0$). In the left panels, the
SNIa is obtained with the SALT-II light curve fit, whereas, in the
right panels, it is given with the MLCS2k2 one. The blue dotted,
yellow dot-dashed, green dashed and pink solid lines show the
results from SDSS-II SNIa, BAO+CMB, SDSS-II SNIa+BAO, and SDSS-II
SNIa+BAO+CMB, respectively. From this figure, it is interesting to
see that,  for all three different parametrization forms, the
$\Lambda$CDM model is consistent with the SNIa, SNIa+BAO and
SNIa+BAO+CMB at the $1\sigma$ confidence level when the SALT-II fit
is used. This consistency is, however, broken at the $2\sigma$
confidence level when the fit is changed  to MLCS2k2. Thus, the
consistency of the $\Lambda$CDM model with data depends crucially on
the type of fit used.  This agrees with what was obtained in
Ref.~\cite{Sanchez2009} from the SDSS-II SNIa with the CPL
parametrization. Let us note that this discrepancy between two
analysis methods has been pointed out in Refs.~\cite{Nesseris2005,
Jassal2006},  and it was also found with a simple $w$CDM model in
the original SDSS-II paper~\cite{Kessler}. Furthermore, we find that
the SDSS-II SNIa with SALT-II fit is consistent with BAO+CMB very
well. However,  when the MLCS2k2 fit is used, there exists a tension
between SNIa and CMB+BAO at the $1\sigma$ confidence level.
Actually, this tension also exists between other SNIa sets, Gold,
for an example, and CMB+BAO~\cite{Jassal2005}.

\begin{figure}[htbp]
 \centering
\includegraphics[width=0.50\textwidth, height=0.40\textwidth]{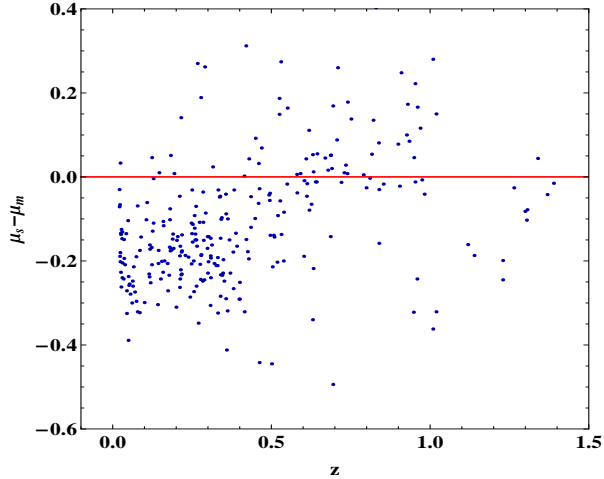}
 \caption{\label{Fig1}  The difference in the distance modulus obtained with two fits for each supernova.
The subscripts $s$ and $m$ represent the  SALT-II fit and  the
MLCS2k2 fit respectively.  }
 \end{figure}

Fig.~(\ref{Fig3}) shows the evolutionary curves of the deceleration
parameter $q(z)$ reconstructed from the SDSS-II SNIa (SALT-II and
MLCS2k2)+BAO and the SDSS-II SNIa+BAO+CMB for three parametrization
forms.  The gray regions and the regions between the two dashed
lines represent the $1\sigma$ confidence level for $q(z)$ obtained
from SNIa+BAO+CMB and SNIa+BAO, respectively. The dot-dashed lines
show the best fit curves of the spatially flat $\Lambda$CDM model.
 The left panels in Fig.~(\ref{Fig3}) reveal that, when the
SALT-II fit is considered, independent of the parametrization forms,
both SNIa+BAO and SNIa+BAO+CMB support an accelerating cosmic
expansion and the acceleration seems to be speeding up. This is
different from the results obtained in Refs.~\cite{slowing, gong}
where it was found that the Constitution SNIa+BAO favor a
slowing-down of the cosmic acceleration at the low redshifts.
However, the right panels  show that, for the MLCS2k2 fit, the
best-fit results obtained from SNIa+BAO+CMB for the CPL and UIS
parametrization forms  favor a slowing-down of the cosmic
acceleration. While, when the JBP parametrization form is used, the
results obtained from SNIa+BAO+CMB  change markedly and an
increasing cosmic acceleration is favored. If the CMB data is not
included, the results are consistent for three parametrization forms
and observations favor an increase of the cosmic acceleration. In
addition, we also find that the SALT-II SDSS-II SNIa gives a
best-fit current acceleration which is larger than that from the
$\Lambda$CDM, whereas the MLCS2k2 SDSS-II SNIa gives a one which is
less. Thus, different light curve fits of SNIa may yield completely
different behavior of dark energy and the parametrization forms also
matter. For the purpose of unfolding the uncertainty in supernova
light curves, we plot the difference of the distance modulus
obtained with two fits for each supernova in Fig~(\ref{Fig1}). From
this figure, it is easy to see that for most low redshift supernova
data points ($z$ is about less than $0.5$) their distance moduli
from the SALT-II fit are smaller than those from the MLCS2k2 fit,
which apparently lead to  different results for the cosmic evolution
from SNIa.

\begin{figure}[htbp]
 \centering
\includegraphics[width=0.45\textwidth, height=0.38\textwidth]{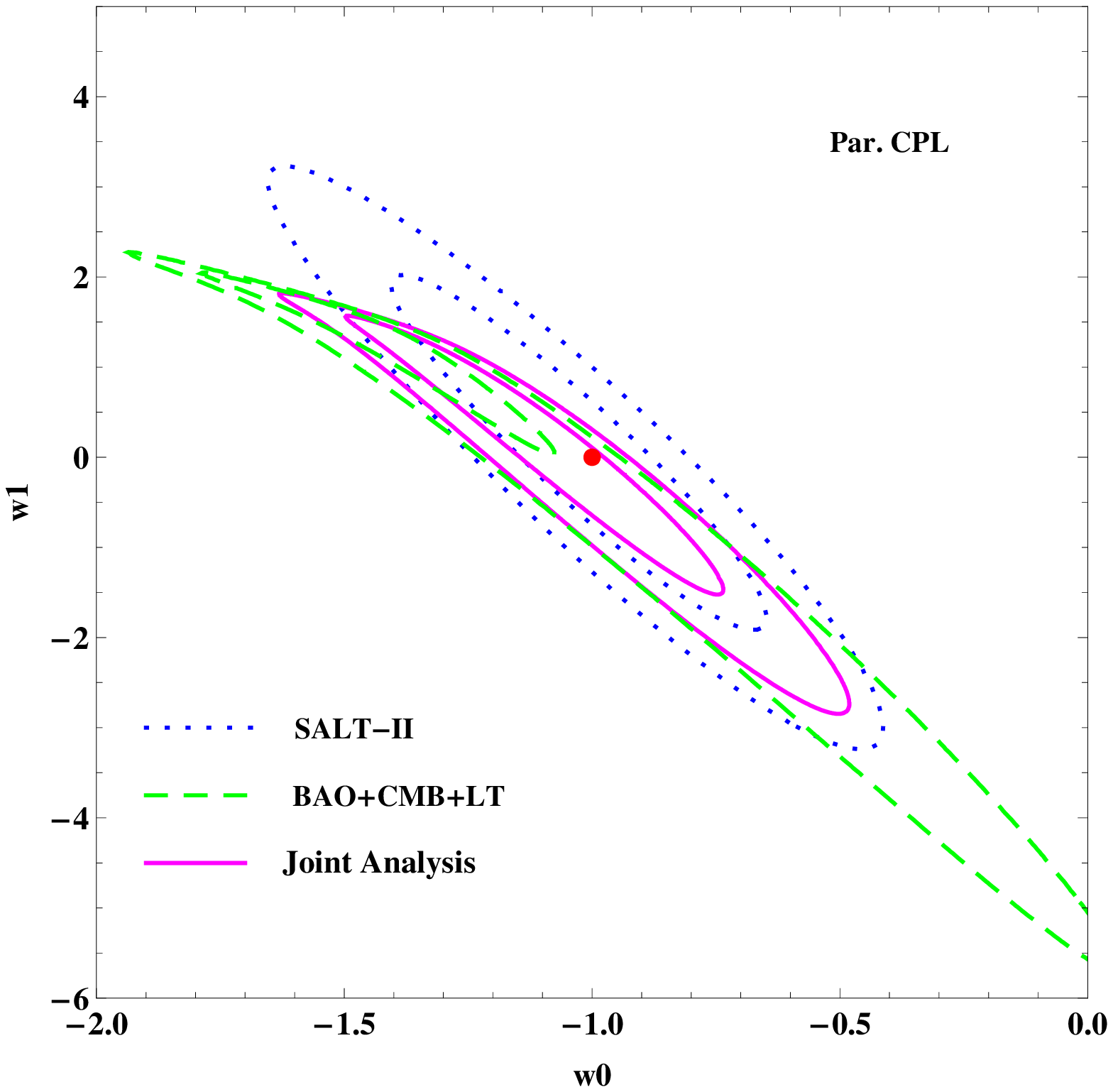}\includegraphics[width=0.45\textwidth, height=0.38\textwidth]{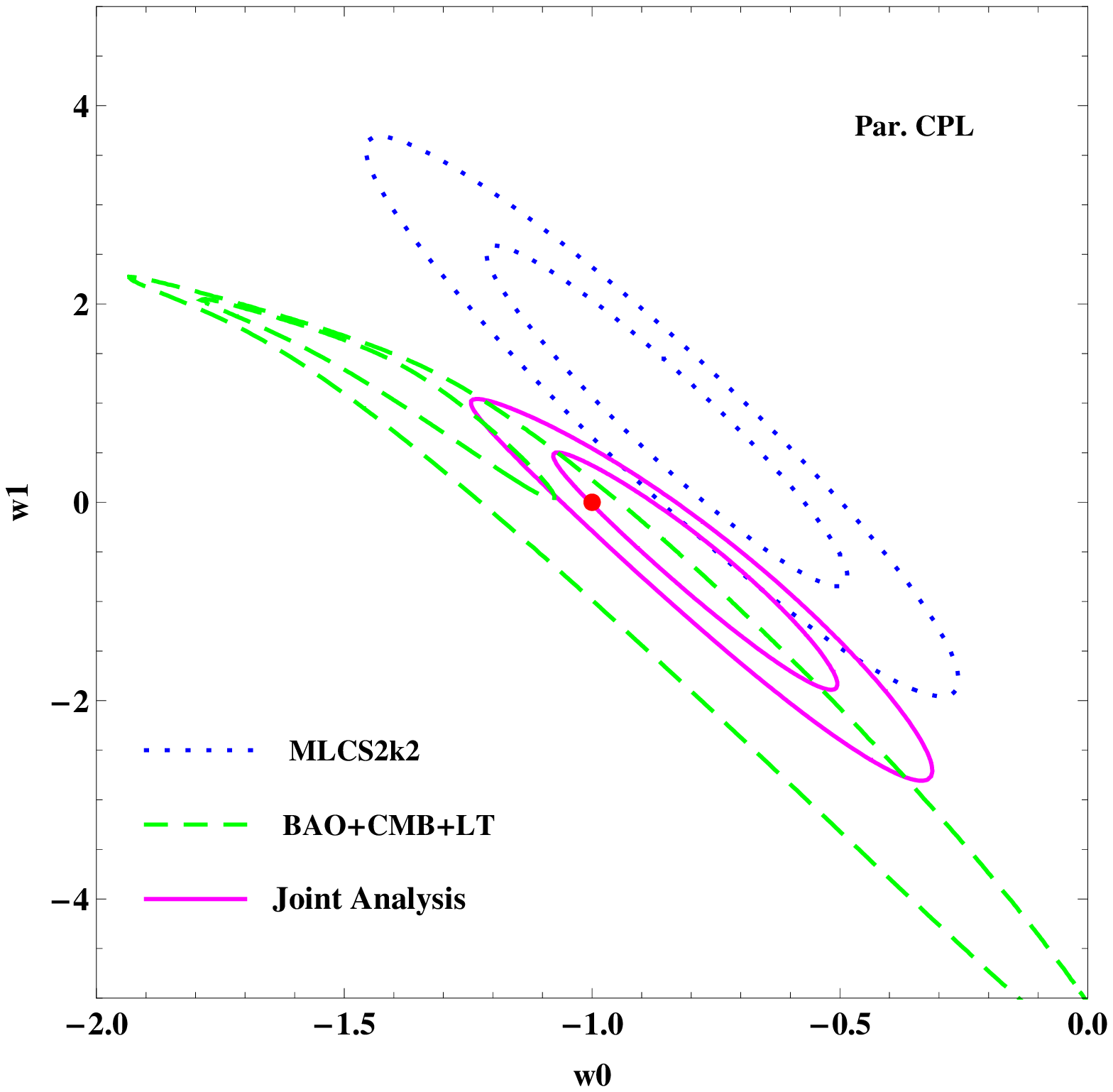}
\includegraphics[width=0.45\textwidth, height=0.38\textwidth]{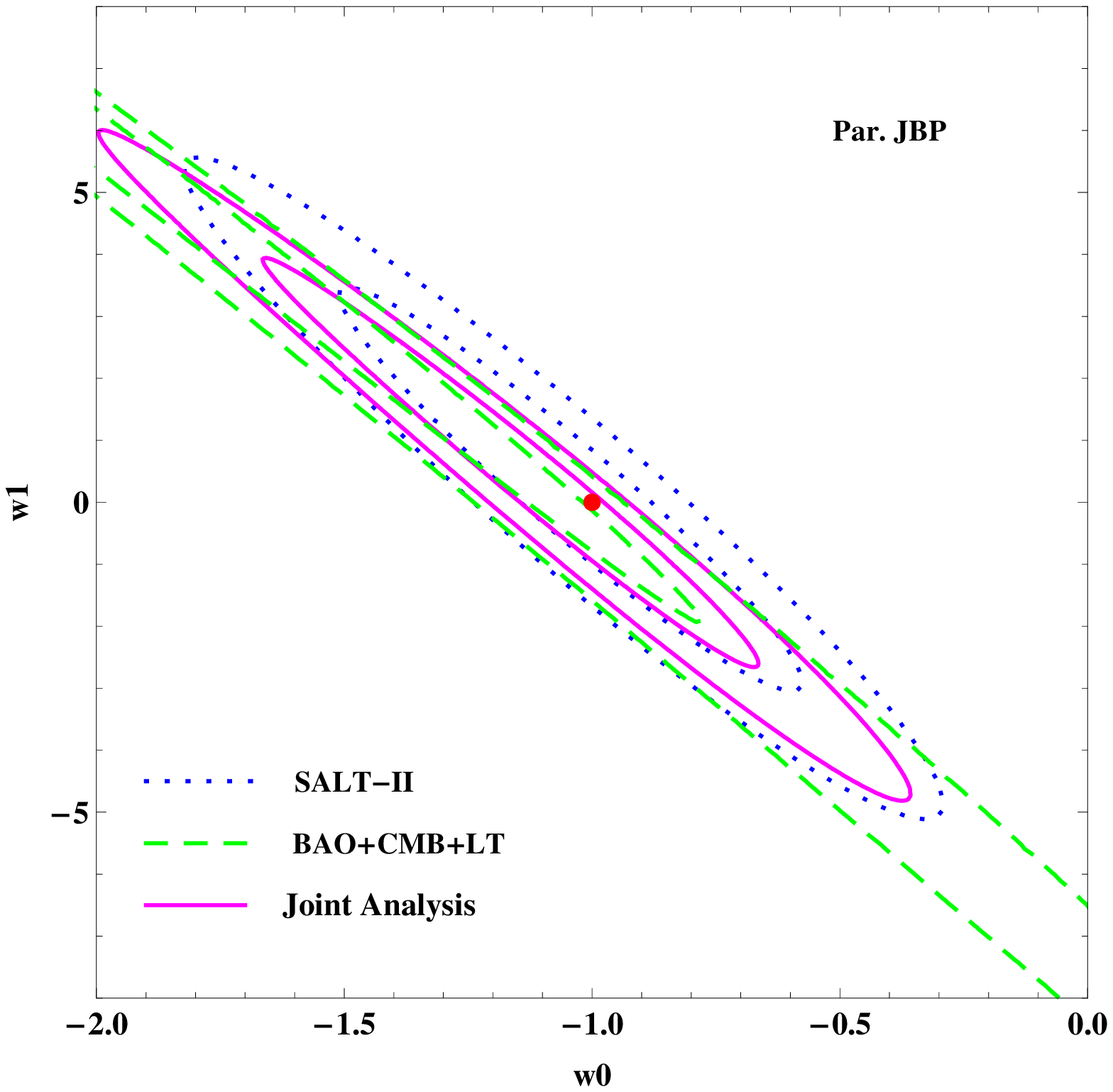}\includegraphics[width=0.45\textwidth, height=0.38\textwidth]{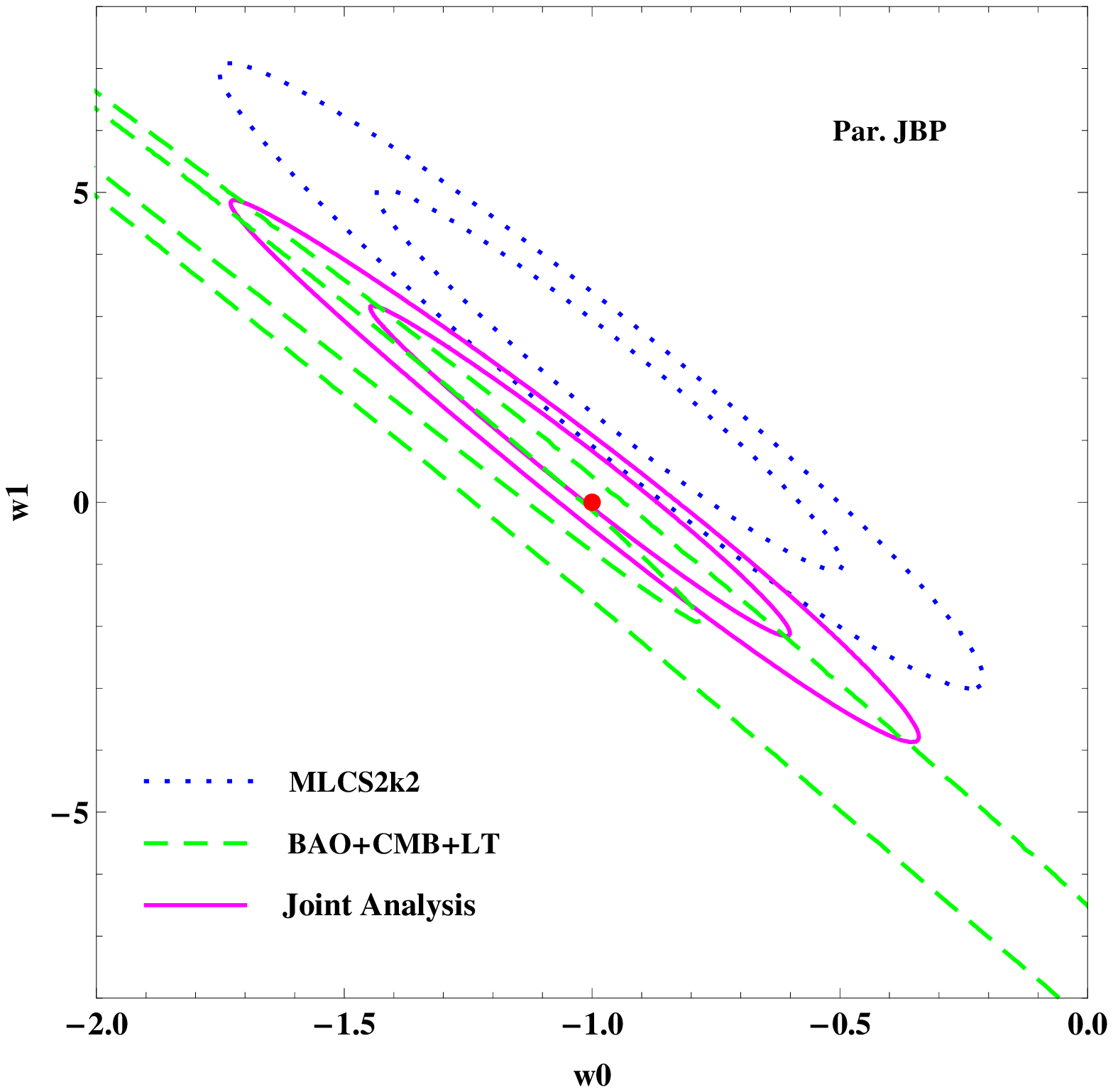}
\includegraphics[width=0.45\textwidth, height=0.38\textwidth]{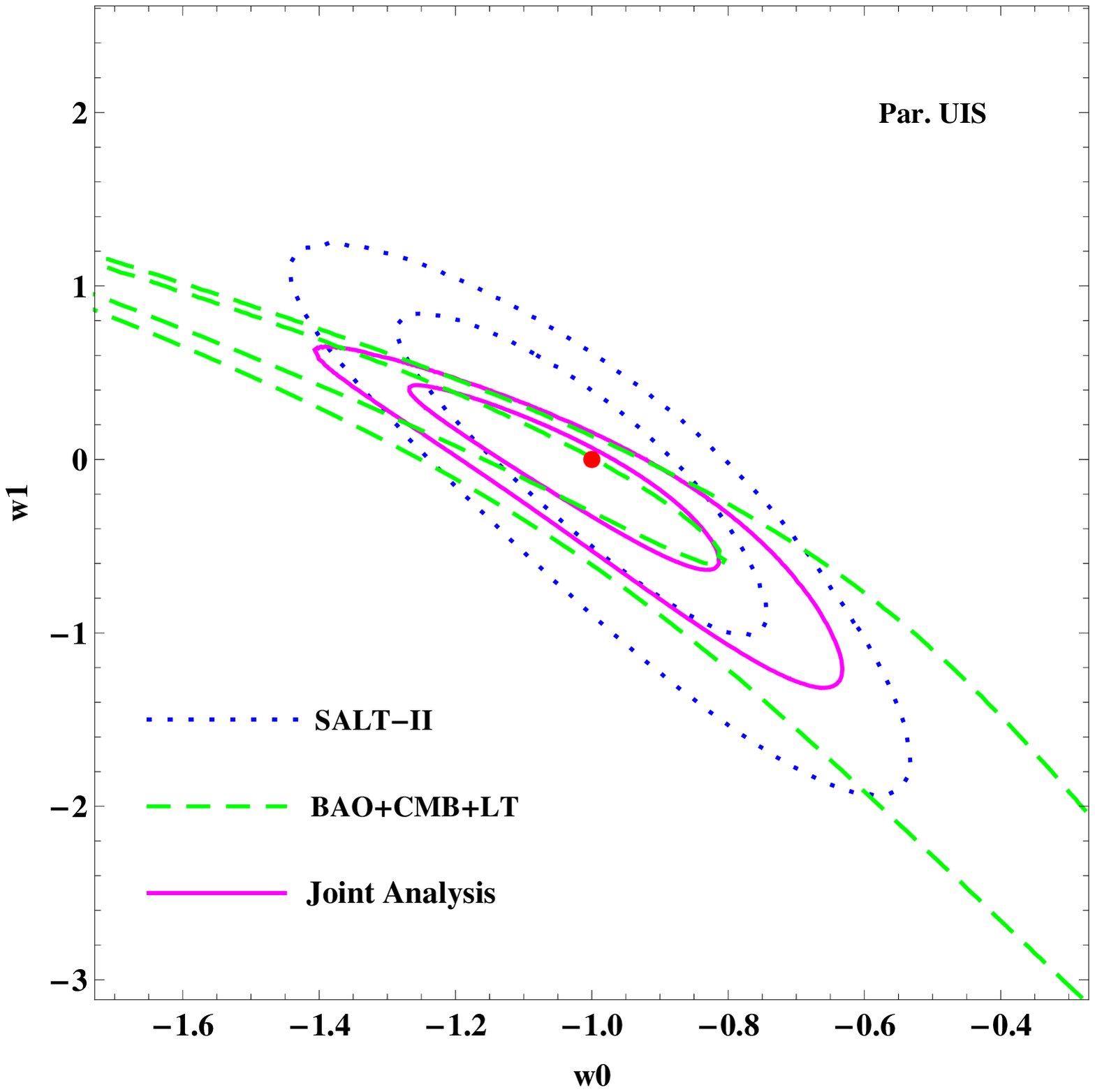}\includegraphics[width=0.45\textwidth, height=0.38\textwidth]{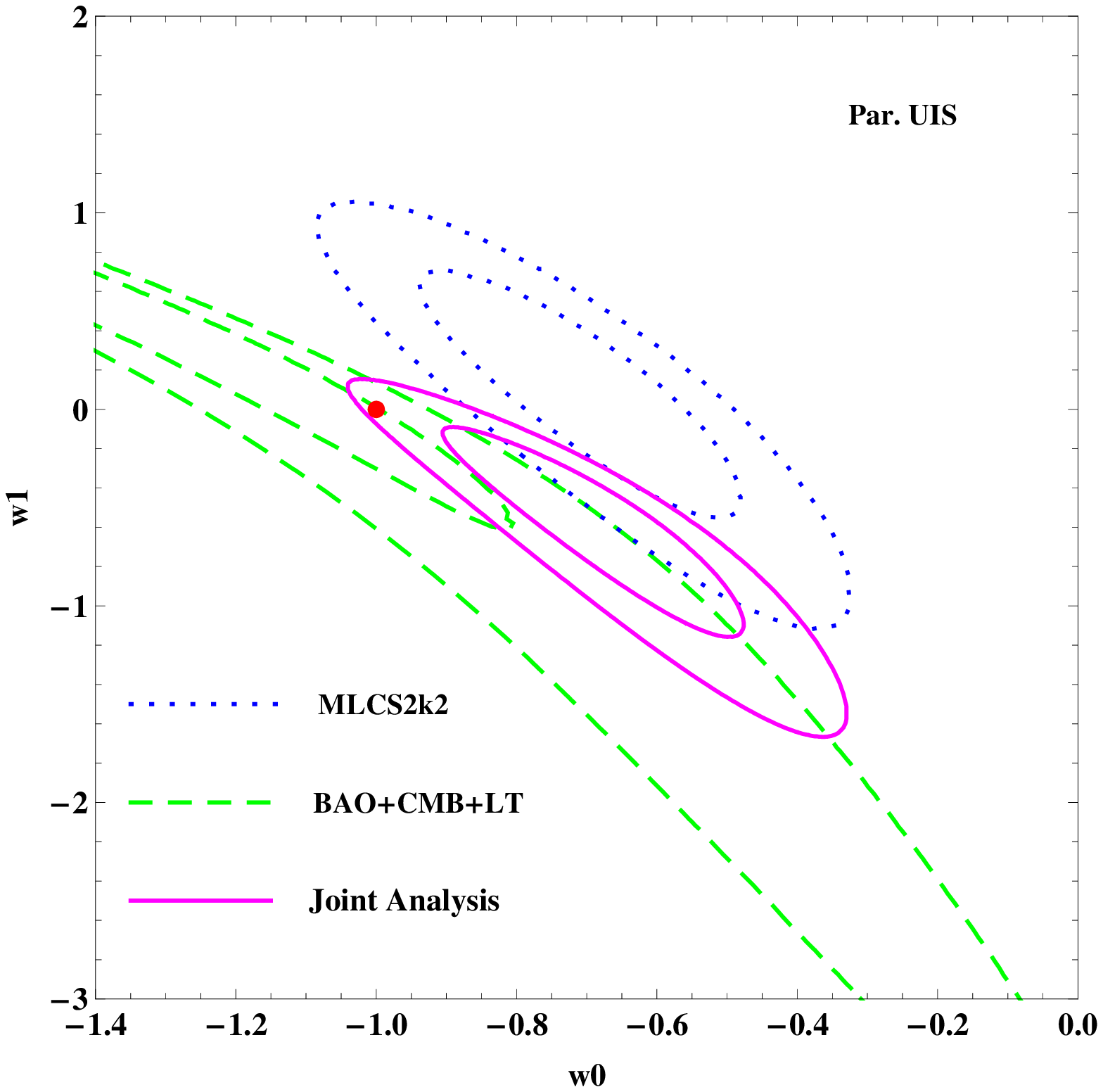}
 \caption{\label{Fig5} The $68.3\%$ and $95\%$ confidence level regions
   for $w_0$ versus $w_1$. The blue dotted, green dashed and
pink solid lines represent the results obtained from SNIa,
BAO+CMB+LT and SNIa+BAO+CMB+LT, respectively.  The left panels show
the results obtained from SNIa with SALT fit, while the right panels
are the results from SNIa with MLCS2k2 fit.
   The red point at $w_0=-1$, $w_1=0$
   represents the spatially flat $\Lambda$CDM model.}
 \end{figure}

\begin{figure}[htbp]
 \centering
\includegraphics[width=0.45\textwidth, height=0.38\textwidth]{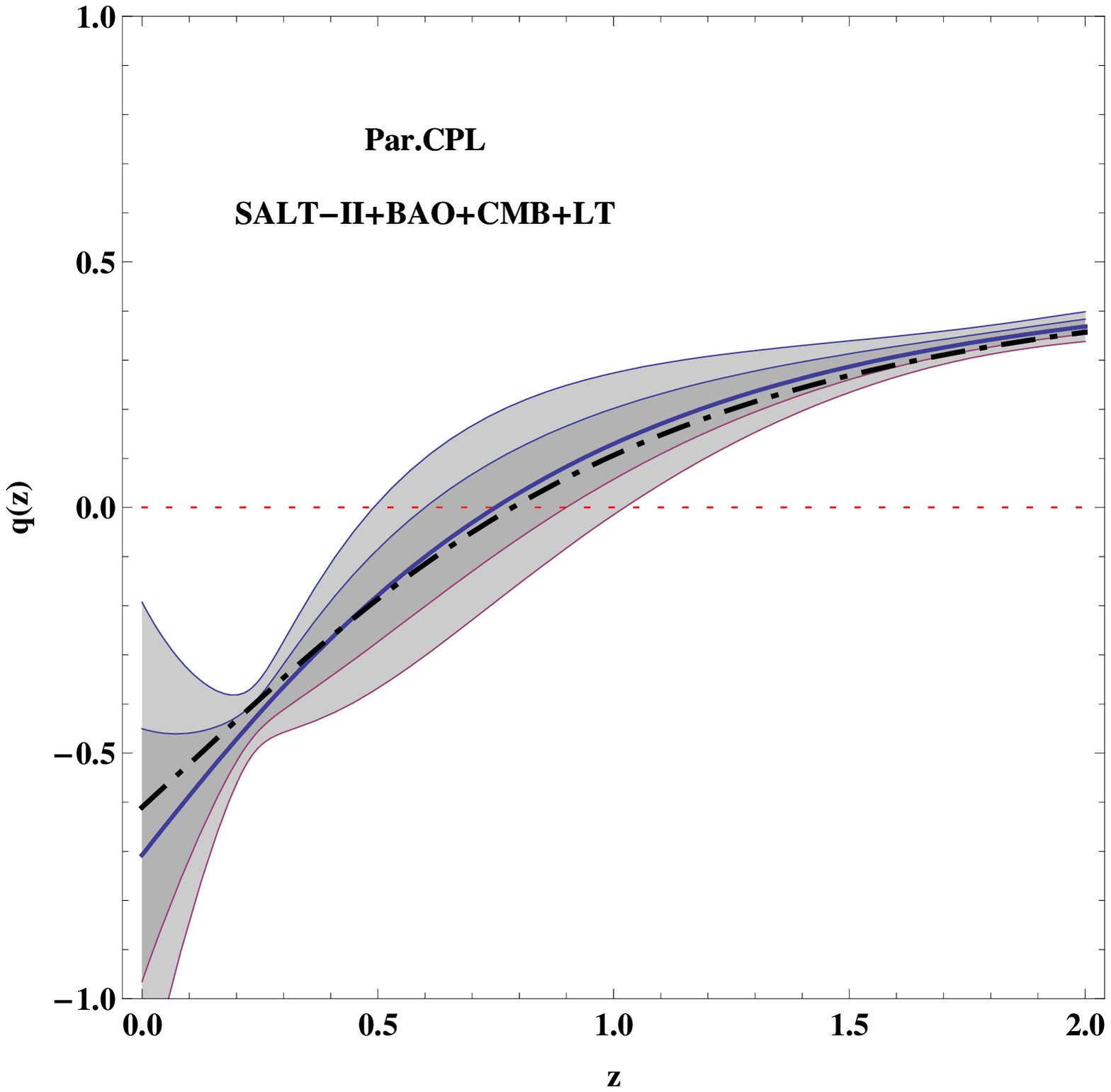}\includegraphics[width=0.45\textwidth, height=0.38\textwidth]{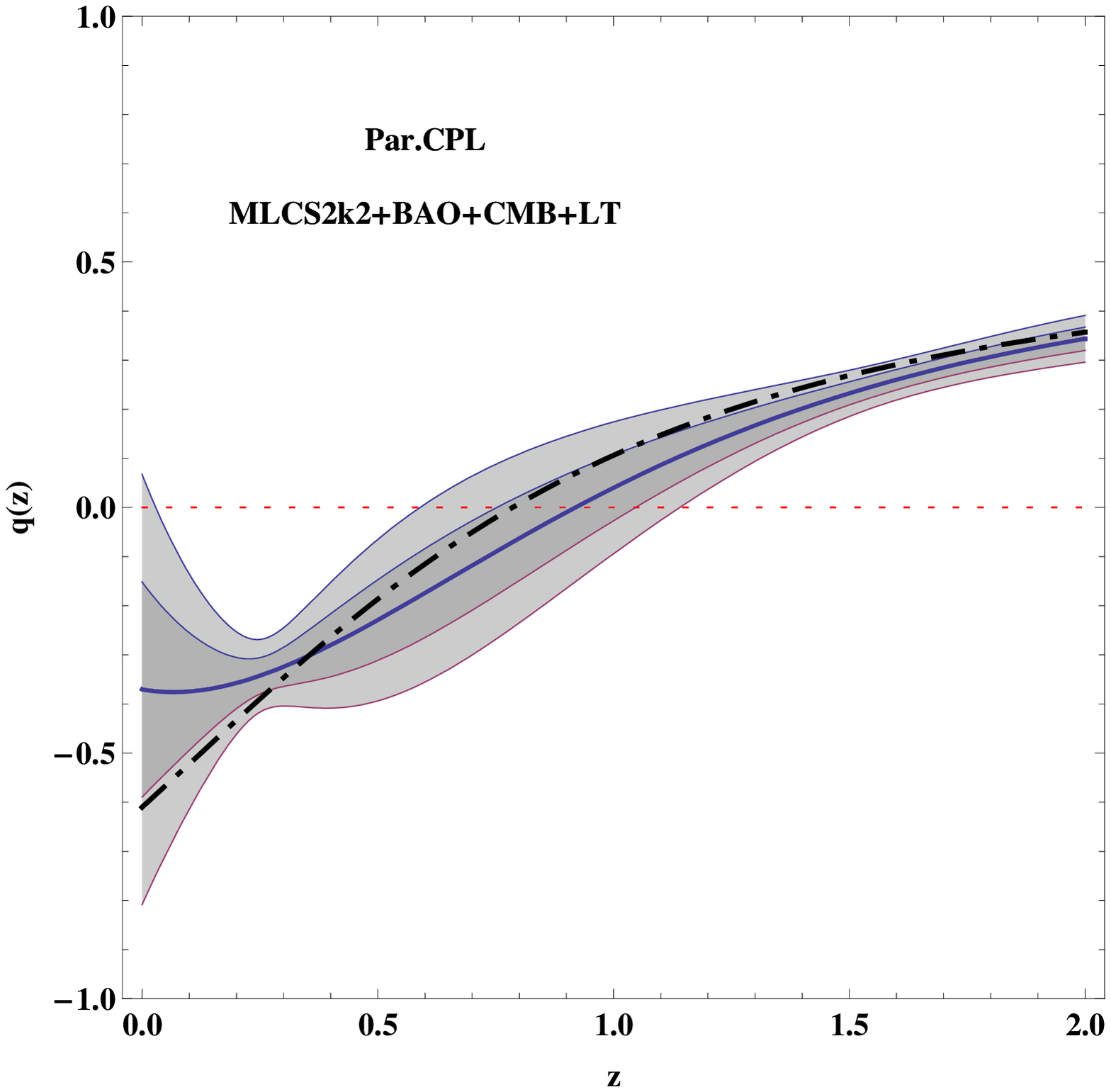}
\includegraphics[width=0.45\textwidth, height=0.38\textwidth]{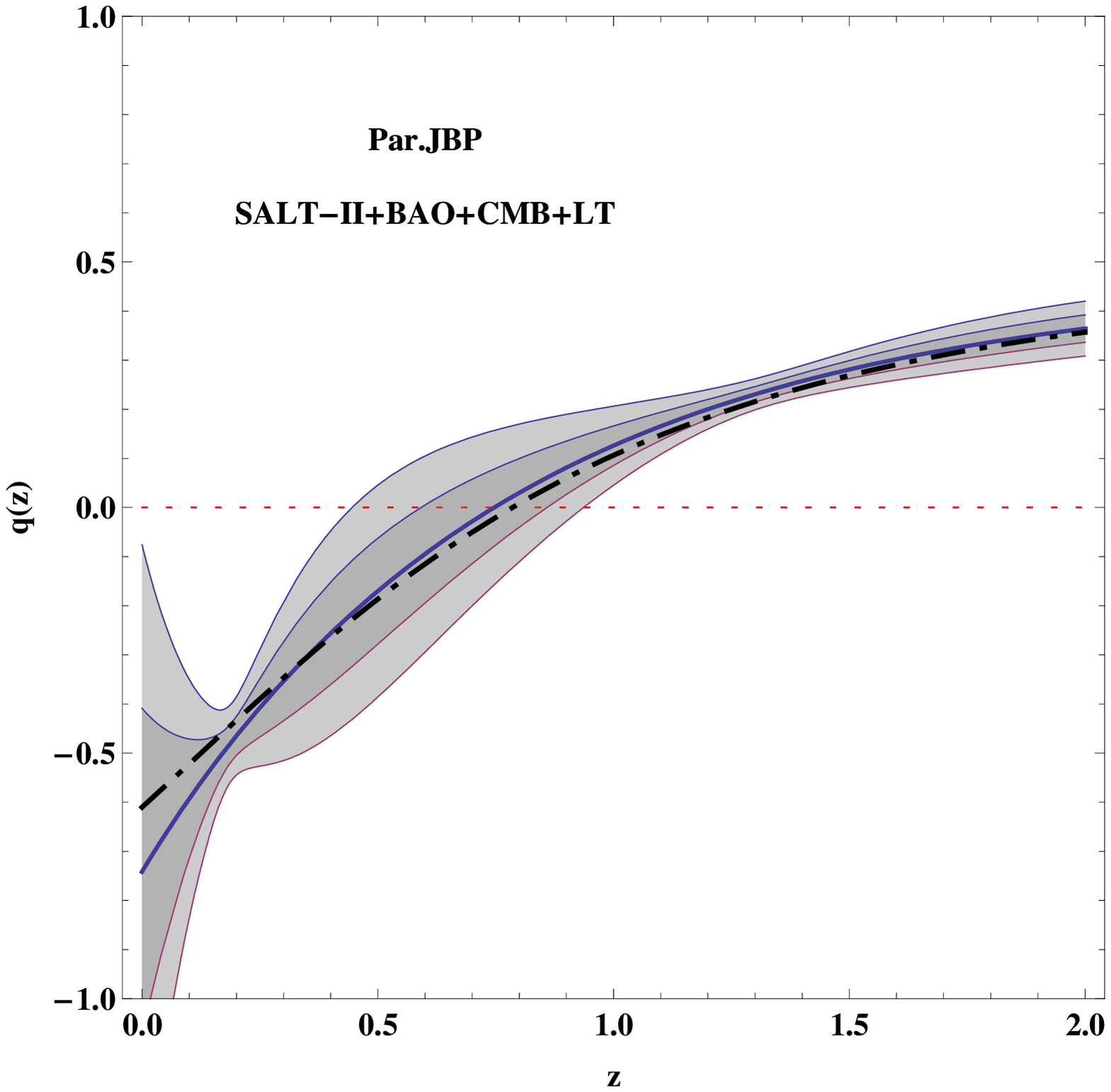}\includegraphics[width=0.45\textwidth, height=0.38\textwidth]{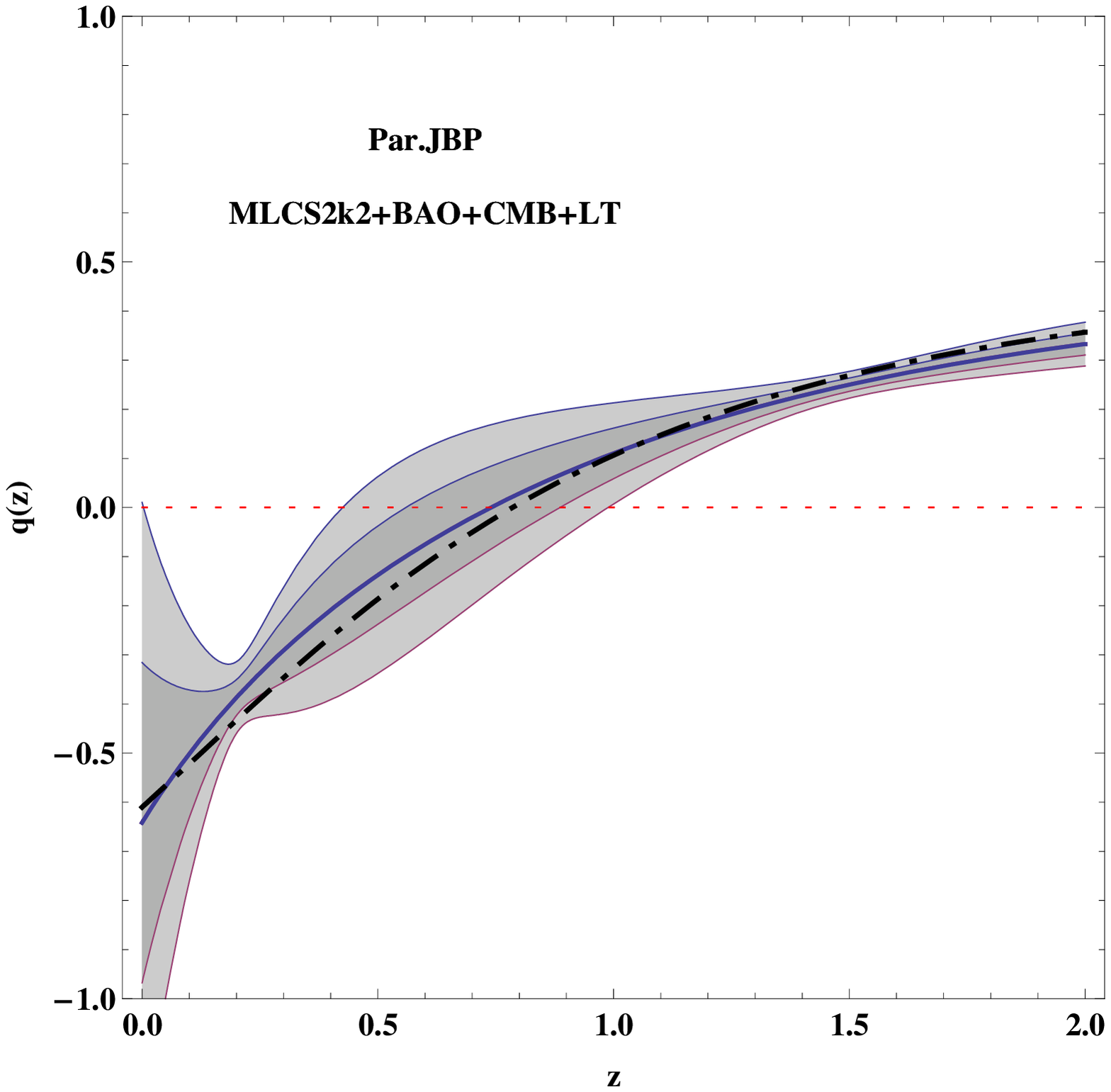}
\includegraphics[width=0.45\textwidth, height=0.38\textwidth]{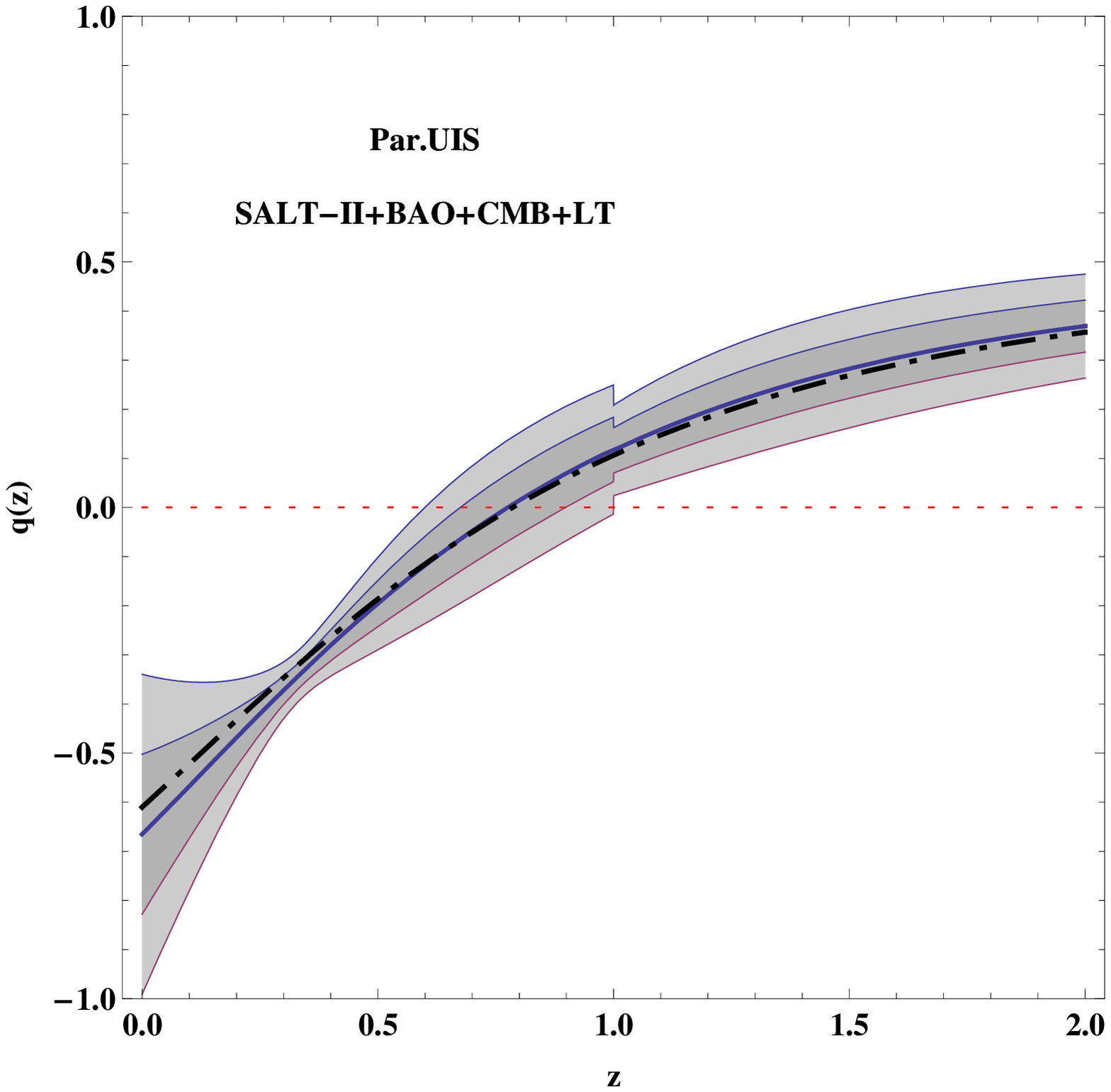}\includegraphics[width=0.45\textwidth, height=0.38\textwidth]{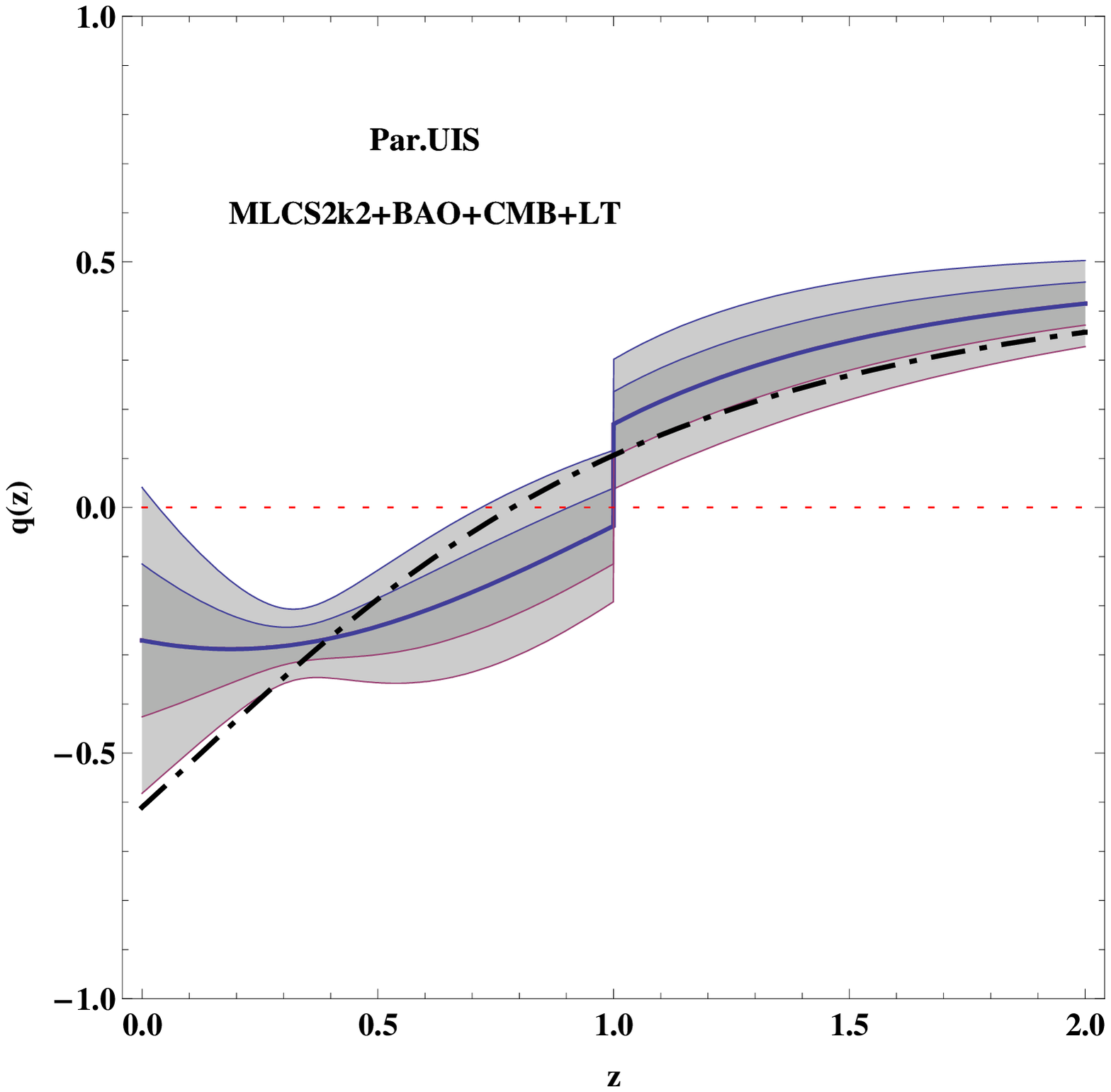}
\caption{\label{Fig6}  The gray regions show the evolutionary
behaviors of $q(z)$ at the $68.3\%$ and $95\%$ confidence levels
obtained from SNIa+BAO+CMB+LT. The dot-dashed lines represent the
best-fit spatially flat $\Lambda$CDM model.  The left panels show
the results obtained from SNIa with SALT fit, while the right panels
are the results from SNIa with MLCS2k2 fit.}
 \end{figure}

\begin{figure}[htbp]
 \centering
\includegraphics[width=0.45\textwidth, height=0.38\textwidth]{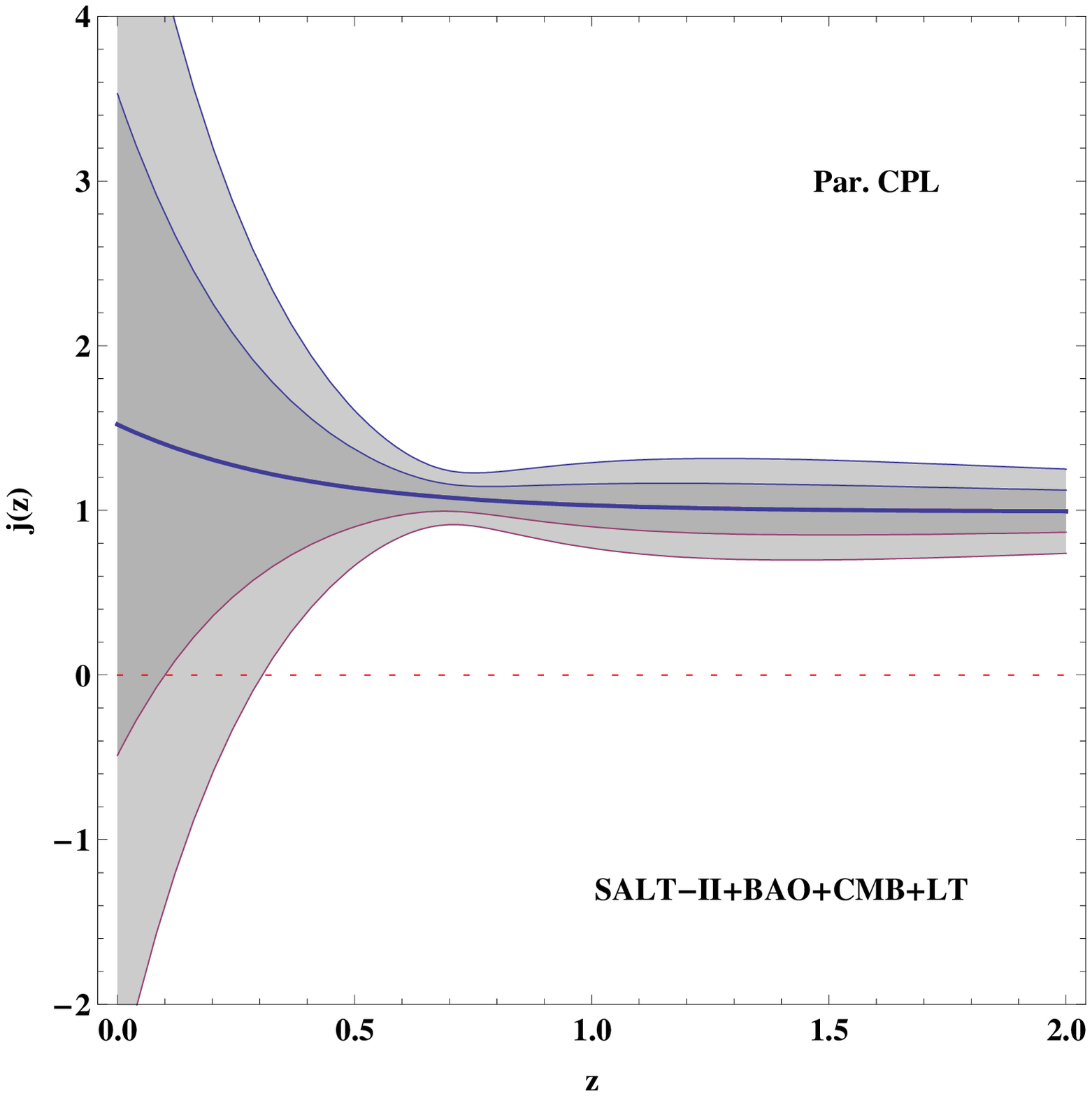}\includegraphics[width=0.45\textwidth, height=0.38\textwidth]{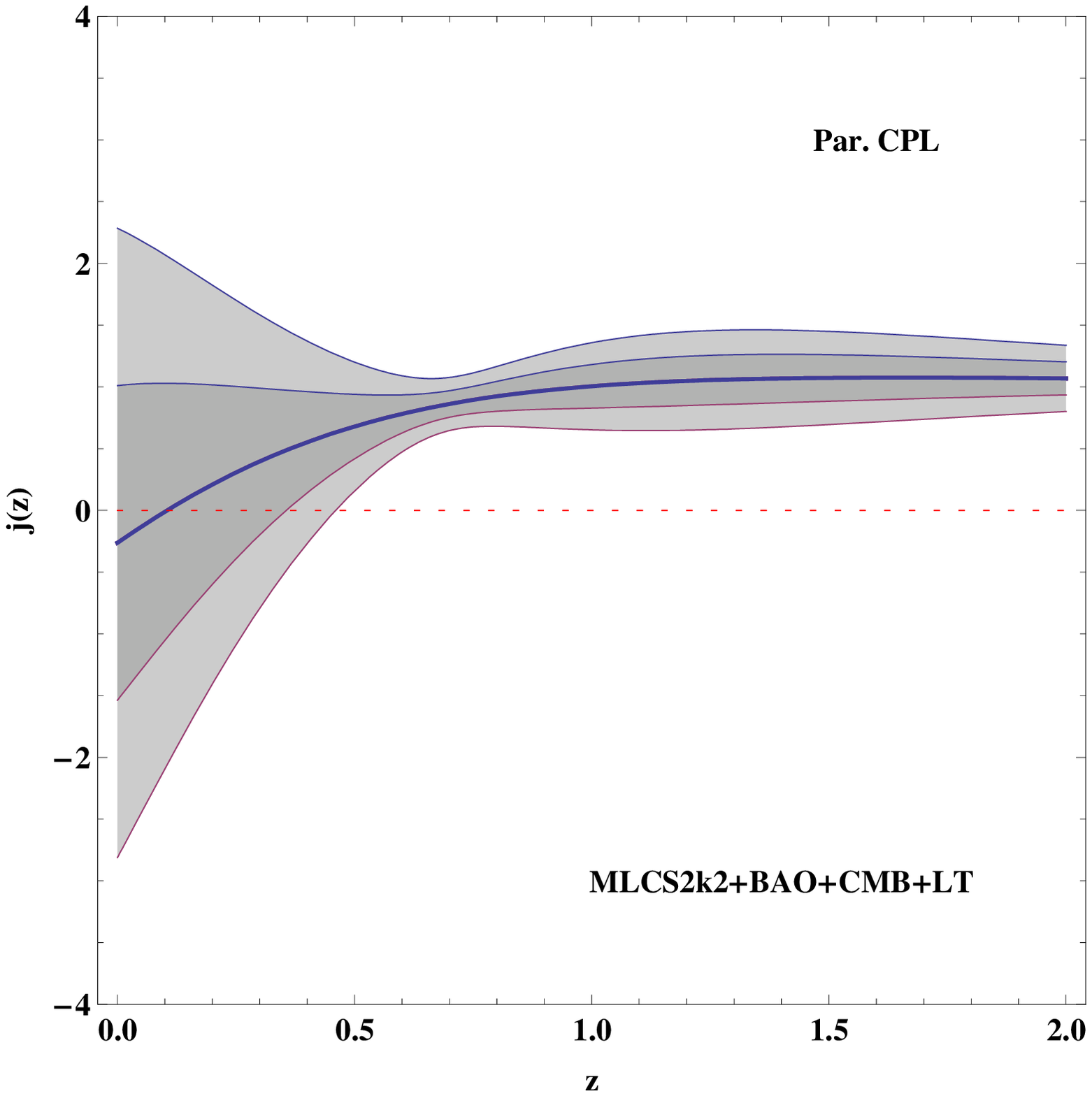}
\includegraphics[width=0.45\textwidth, height=0.38\textwidth]{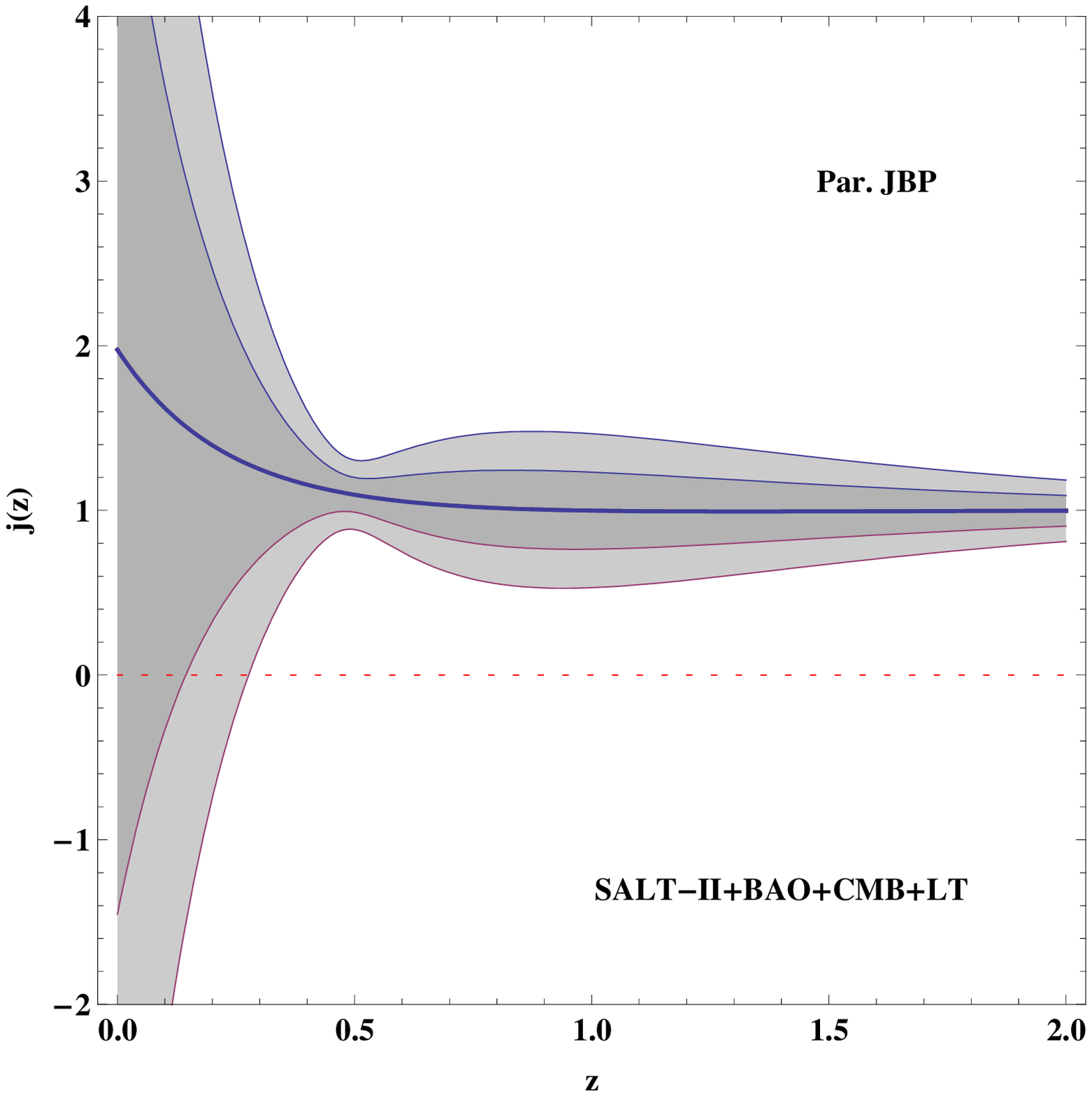}\includegraphics[width=0.45\textwidth, height=0.38\textwidth]{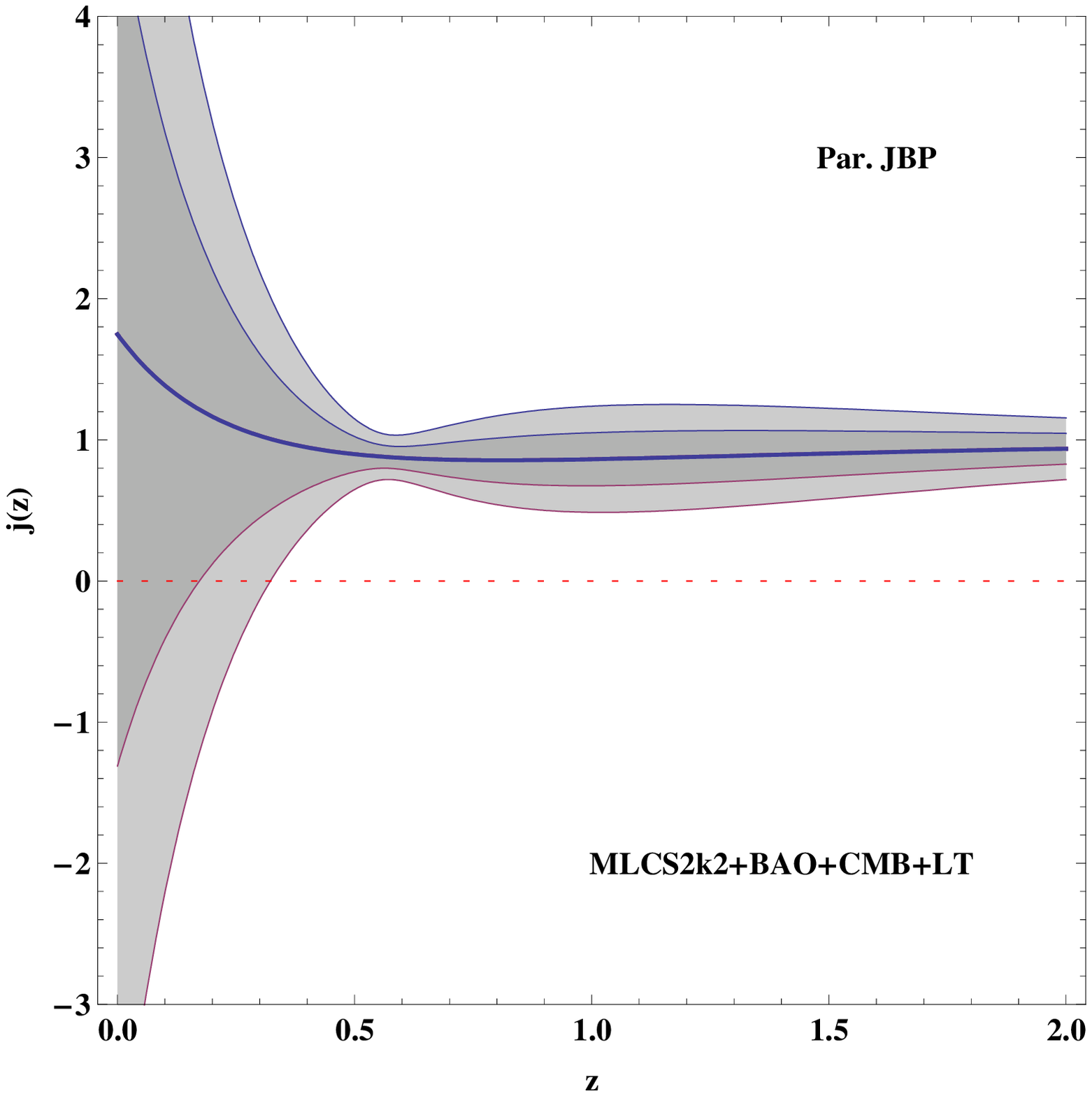}
\includegraphics[width=0.45\textwidth, height=0.38\textwidth]{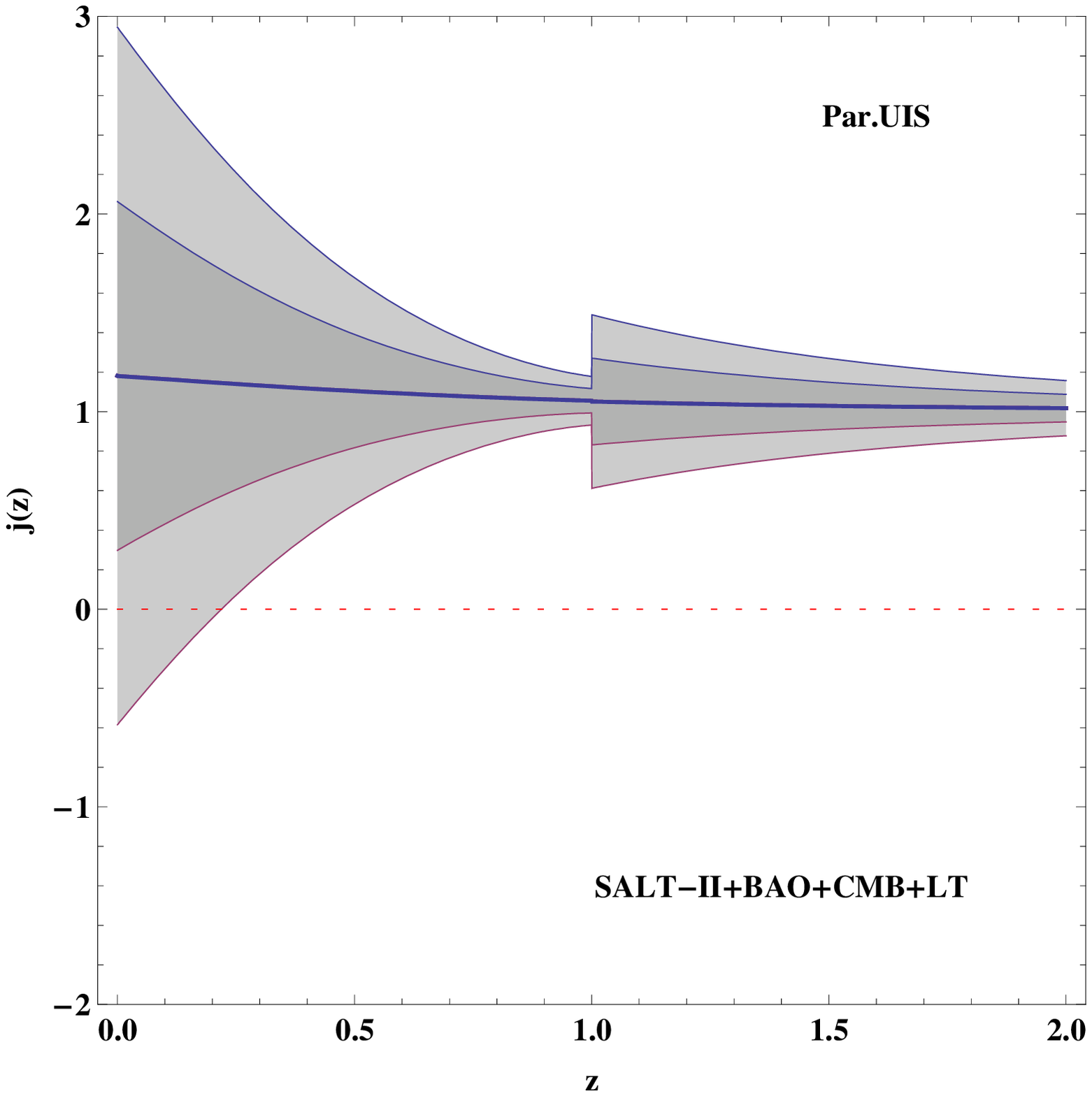}\includegraphics[width=0.45\textwidth, height=0.38\textwidth]{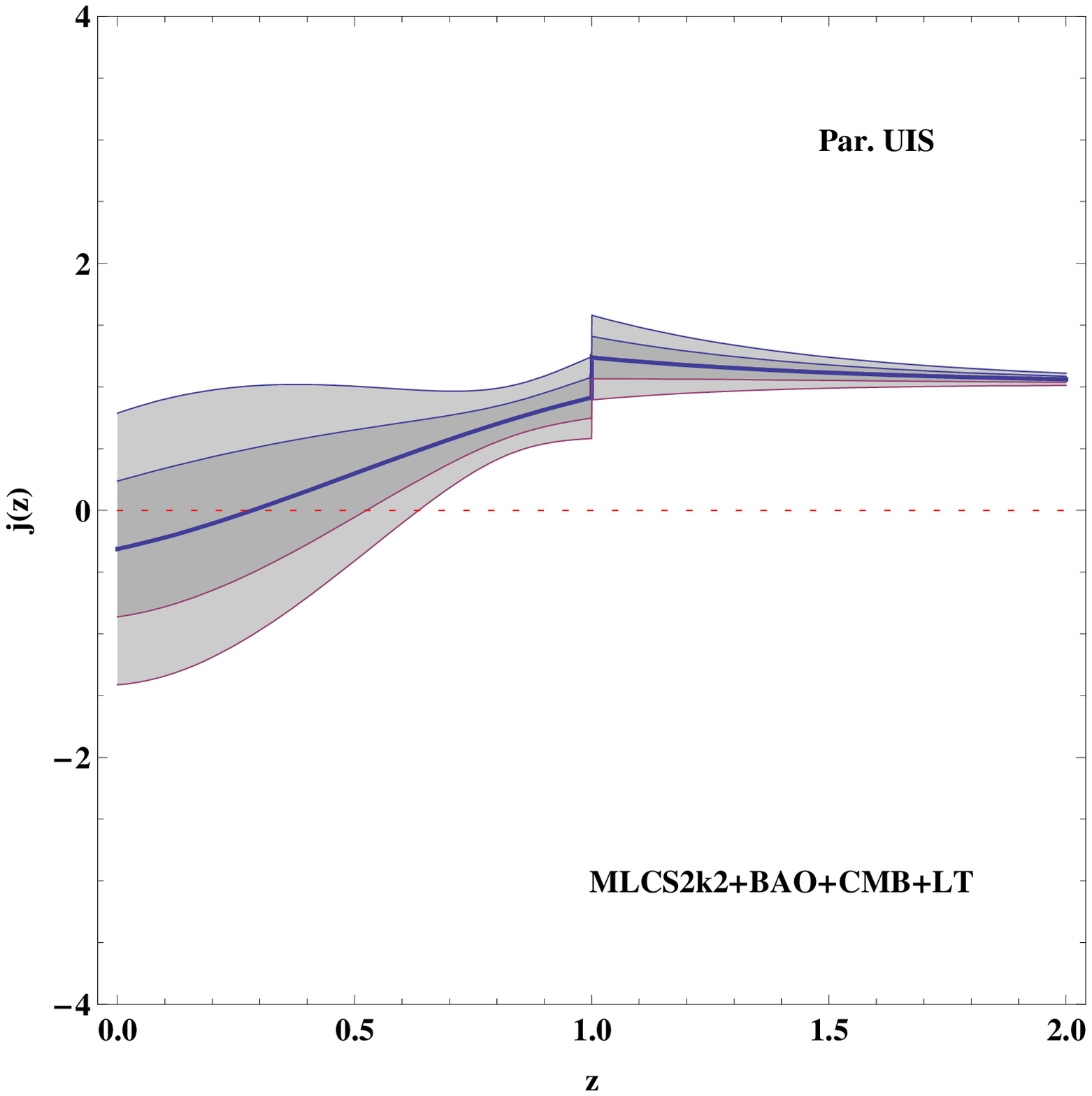}
 \caption{\label{Fig7}   The gray regions show the evolutionary behaviors of $j(z)$ at the
 $68.3\%$ and $95\%$
 confidence levels obtained from SNIa+BAO+CMB+LT.  The left
panels show the results obtained from SNIa with SALT fit, while the
right panels are the results from SNIa with MLCS2k2 fit.}
 \end{figure}

Since the LT has the advantage of avoiding the biases existing in
data obtained from the cosmic distance ladder method,  we add it in
our discussion to obtain more information on the evolutional
behavior of dark energy. The results are shown in Figs.~(\ref{Fig5},
 \ref{Fig6}, \ref{Fig7}).
Fig~(\ref{Fig5}) gives the allowed region of  the model parameters
$w_0$ and $w_1$ at the $68.3\%$ and $95.4\%$ confidence levels. The
dotted, dashed and solid lines represent the results  from SNIa,
BAO+CMB+LT and SNIa+BAO+CMB+LT, respectively. Comparing the right
panels of Fig.~(\ref{Fig2}) and Fig.~(\ref{Fig5}), one can see that
when LT data is added, the tension between the MLCS2k2 SDSS-II SNIa
and other data sets  becomes more severe  and it exists at the
$2\sigma$ confidence level. However, LT data render the results to
be more consistent with the $\Lambda$CDM, since MLCS2k2 SDSS-II
SNIa+BAO+CMB+LT allow the $\Lambda$CDM, but  MLCS2k2 SDSS-II
SNIa+BAO+CMB  rule out it at the $2\sigma$ confidence level.

Fig.~(\ref{Fig6}) gives the evolutionary curve of $q(z)$
reconstructed from SNIa+BAO+CMB+LT.    The gray regions show
$68.3\%$ and $95.4\%$ confidence levels, the thick solid lines are
the best fit results, and the dot-dashed lines indicate the
spatially flat $\Lambda$CDM. We find that the results are similar to
that obtained from  SDSS-II SNIa+BAO+CMB. Only in the cases of the
CPL and UIS  parametrizations  and the SDSS-II SNIa with MLCS2k2
fit, do SNIa+BAO+CMB+LT favor that the cosmic acceleration is
slowing down. In all other cases,  an increasing of the cosmic
acceleration is favored. That is to say, the inclusion of the LT
data does not change the evolutionary behavior of $q(z)$ markedly.

 To further confirm quantitatively the behavior of $q(z)$,  we plot,
  in Fig.~(\ref{Fig7}),
the evolutionary curves of the jerk parameter $j(z)$, which is
proportional to the derivative of $q(z)$. From the left panel of
this figure, one can see that SALT-II SDSS-II SNIa+BAO+CMB+LT favor
an increasing cosmic acceleration at the present since  the value of
$j(0)$ is largely in the region of $> 0$, but they cannot rule out
the possibility of a decrease of the cosmic acceleration since
$j(0)<0$ is still allowed at the $2\sigma$ confidence level. The
right panel of Fig.~(\ref{Fig7}) shows that although, for the CPL
and the UIS parametrizations, observations favor a slowing down of
the cosmic acceleration,  they still allow a currently increasing
cosmic acceleration.  For the JBP parametrization,  similar results
from SDSS-II SNIa with SALT-II and MLCS2k2 fits are obtained, and
observations always favor an increasing cosmic acceleration.

\section{CONCLUSION}\label{sec4}
 We focus, in this paper, on analyzing the property of dark energy  by
reconstructing  the deceleration parameter $q(z)$ from the SDSS-II
SNIa with two different light curve fits (SALT-II and MLCS2k2) along
with BAO, CMB and LT. Three different parametrization forms of the
EOS of dark energy (CPL, JBP and UIS) are investigated. We find
that, when the SALT-II light curve fit is considered, independent of
the parametrization forms  SNIa+BAO, SNIa+BAO+CMB as well as
SNIa+BAO+CMB+LT favor an accelerating cosmic expansion which is
currently speeding up. We also find that there is an tension between
SNIa and other data sets (CMB+BAO or CMB+BAO+LT). These differ from
what was obtained from the Constitution SNIa+BAO in
Refs.~\cite{slowing,gong}, where it was found that the cosmic
acceleration is probably slowing down now. However, when  the
MLCS2k2 light curve fit  is considered, the results are dependent on
the parametrization forms. For the CPL and UIS parametrization
forms, the best fit results obtained from SNIa+BAO+CMB and
SNIa+BAO+CMB+LT show that a slowing-down acceleration of the cosmic
expansion is favored, and  this, however, does not occur  for
SNIa+BAO.  For the JBP parametrization,  an increasing cosmic
acceleration is always favored. Finally,  from the evolutionary
curves of $j(z)$, we find that  both  an increasing and a decreasing
of the current cosmic acceleration are allowed by the
SNIa+BAO+CMB+LT at the 2$\sigma$ confidence level.

Thus, the evolutional behavior of dark energy reconstructed and the
issue of whether the cosmic acceleration is slowing down or even
speeding up is highly dependent upon the  SNIa data sets, the light
curve fitting method of the SNIa, and the parametrization forms of
the equation of state. In order to have a definite answer, we must
wait for data with more precision and search for the more reliable
and efficient methods to analyze these data.

\section*{ACKNOWLEDGEMENTS}
This work was supported in part by the National Natural Science
Foundation of China under Grants Nos. 10775050, 10705055, 10935013
and 11075083,  Zhejiang Provincial Natural Science Foundation of
China under Grant No. Z6100077, the SRFDP under Grant No.
20070542002, the FANEDD under Grant No. 200922, the National Basic
Research Program of China under Grant No. 2010CB832803, the NCET
under Grant No. 09-0144, and the PCSIRT under Grant No. IRT0964.


\begin{thebibliography}{99}
\bibitem{expansion1}
        A. G.Riess, {\it et al.}, Astron. J. 116, 1009 (1998).
\bibitem{expansion2}
        S. Perlmutter, {\it et al.}, Astrophy. J. 517, 565 (1999).
\bibitem{observation1}
        G. Hinshaw, {\it et al.}, Astrophys. J. Suppl. Ser. 180, 225 (2009);
        E. Kommatsu, {\it et al.}, Astrophys. J. Suppl. Ser. 180, 230 (2009).
\bibitem{observation2}
        A. Vikhlinin, {\it et al.}, Astrophys. J. 692, 1060 (2009).
\bibitem{observation3}
        E. Rozo, {\it et al.}, Astrophys. J. 708, 645 (2010).
\bibitem{cross1}
        U. Alam, V. Sahni, T. D. Saini, A. A. Starobinsky, Mon. Not. Roy. Astron. Soc. 354, 275 (2004);
     U. Alam, V. Sahni, A. A. Starobinsky, J. Cosmol. Astropart. P. 0406, 008 (2004);
      Y. Wang and P. Mukherjee, Astrophys. J. 606, 654 (2004);
      R. Lazkoz, S. Nesseris and L. Perivolaropoulos, J. Cosmol. Astropart. P. 0511, 010 (2005);
      S. Nesseris and L. Perivolaropoulos, J. Cosmol. Astropart. P. 0701, 018 (2007);
      P. Wu and H. Yu, Phys. Lett. B 643, 315 (2006)
\bibitem{cross2}Y. G. Gong and A. Wang, Phys. Rev. D 75, 043520 (2007);
        Y. G. Gong, R. G. Cai, Y. Chen and Z. H. Zhu, J. Cosmol. Astropart. Phys. 01,   019 (2010).
\bibitem{CPL}
        M. Chevallier and D. Polarski, Int. J. Mod. Phys. D 10, 213 (2001);
        E. V. Linder, Phys. Rev. Lett. 90, 091301 (2003).
\bibitem{JBP}
        H. K. Jassal, J. S. Bagla, and T. Padmanabhan, Mon. Not. R. Astron. Soc. 356, L11 (2005);
        T. Roy Choudhary and T. Padmanabhan, Astron. Astrophys. 429, 807 (2005).
\bibitem{UIS}
        A. Upadhye, M. Ishak, ahd P. Steinhardt, Phys. Rev. D 72, 063501 (2005).
\bibitem{diagnostic}
        V. Sahni, A. Shafieloo and A. A. Starobinsky, Phys. Rev. D 78, 103502 (2008).
\bibitem{constitution}
        M. Hicken, {\it et al.}, Asrophys. J. 700, 1097 (2009).
\bibitem{bao1}
        W. J. Percival, {\it et al.}, Mon. Not. R. Astron. Soc. 381, 1053 (2007).
\bibitem{bao2}
        B. A. Reid, arXiv: 0907.1659; W. J. Percival, {\it et al.}, arXiv: 0907.1660.
\bibitem{slowing}
        A. Shafieloo, V. Sahni and A. A. Starobinsky, Phys. Rev. D 80, 101301 (2009).
\bibitem{gong}
        Y. G. Gong, B. Wang, R. G. Cai, J. Cosmol. Astropart. Phys. 04,  019 (2010).
\bibitem{Kessler}
        R. Kessler, {\it et al.}, Astrophys. J. Suppl. Ser. 185, 32 (2009).
\bibitem{Frieman}
        J. A. Frieman, {\it et al.}, Astron. J. 135, 338 (2008).
\bibitem{Sako}
        M. Sako, {\it et al.}, Astron. J.  135, 348 (2008).
\bibitem{Wood-Vasey}
        W. M. Wood-Vasey, {\it et al.},  Astrophys. J.  666, 694 (2007) .
\bibitem{Astier}
        S. Astier, {\it et al.}, A\&A, 447, 31 (2006).
\bibitem{Riess}
        A. G. Riess, {\it et al.}, Astron. J.  116, 1009 (1998).
\bibitem{Jha}
        S. Jha, A. G. Riess, \& R. P. Kirshner, Astrophys. J. 659, 122 (2007).
\bibitem{Simon}
        J. Simon, L. Verde and J. Jimenez, Phys. Rev. D 71, 123001 (2005).
\bibitem{obl}
        S. Capozziello, {\it et al.}, Phys. Rev. D 70, 123501 (2004);
        N. Pires, Z.-H. Zhu and J. S. Alcaniz, Phys. Rev. D 73, 123530 (2006).
\bibitem{modlike}
        M. A. Dantas, J. S. Alcaniz, D. Jain, A.Dev, Astron. Astrophys. 467, 421 (2007).
\bibitem{WMAP7}
        E. Komatsu, {\it et al.}, arXiv:1001.4538v2
\bibitem{Sanchez2009}J. C. Bueno Sanchez, S. Nesseris and L. Perivolaropoulos, arXiv:0908.2636.
\bibitem{Nesseris2005}S. Nesseris and L. Perivolaropoulos, Phys. Rev. D 72, 123519 (2005).
\bibitem{Jassal2006} H. K. Jassal, J. S. Bagla and T. Padmanabhan, Mon. Not. Roy. Astron. Soc. 405, 2639 (2010) [arXiv:
astro-ph/0601389]
\bibitem{Jassal2005}H. K. Jassal, J. S. Bagla and T. Padmanabhan,
Phys. Rev. D 72, 103503 (2005).

\end{thebibliography}
\end{document}